\documentclass[aip,jcp,amsmath,amssymb,reprint,superscriptaddress,longbibliography]{revtex4-2}
\usepackage[dvipdfmx]{graphicx}
\usepackage{bmpsize}
\usepackage{txfonts}
\usepackage{epsf}
\usepackage{bm}
\usepackage{lipsum}
\usepackage{color}
\usepackage{mathrsfs}

\newcommand{\argmin}{\mathop{\rm argmin}\limits}

\begin{document}
\title{Explaining
reaction coordinates of alanine dipeptide
isomerization obtained from deep neural networks using Explainable Artificial
Intelligence (XAI)}

%%%%%%%%%%%%%%%%%%%%%%%%%%%%%%%%%%%%%%%%%%%%%%%%%%%%%%%%%%%%%%%%%%%%%%%%%%%%%%%%%%%%%%%%%%%%%%%%%%%%%%%%%%%%%%%%%%%%%
\author{Takuma Kikutsuji}
\affiliation{Division of Chemical Engineering, Department of Materials Engineering Science, Graduate School of Engineering Science, Osaka University, Toyonaka, Osaka 560-8531, Japan}

\author{Yusuke Mori}
\affiliation{Division of Chemical Engineering, Department of Materials Engineering Science, Graduate School of Engineering Science, Osaka University, Toyonaka, Osaka 560-8531, Japan}

\author{Kei-ichi Okazaki}
\email{keokazaki@ims.ac.jp}
\affiliation{Research Center for Computational Science, Institute for Molecular Science, Okazaki, Aichi 444-8585, Japan}
\affiliation{The Graduate University for Advanced Studies, Okazaki,
Aichi 444-8585, Japan}

\author{Toshifumi Mori}
\email{toshi\_mori@cm.kyushu-u.ac.jp}
\affiliation{Institute for Materials Chemistry and
Engineering, Kyushu University, Kasuga, Fukuoka 816-8580, Japan}
\affiliation{Interdisciplinary Graduate School of Engineering Sciences,
Kyushu University, Kasuga, Fukuoka 816-8580, Japan}

\author{Kang Kim}
\email{kk@cheng.es.osaka-u.ac.jp}
\affiliation{Division of Chemical Engineering, Department of Materials Engineering Science, Graduate School of Engineering Science, Osaka University, Toyonaka, Osaka 560-8531, Japan}

\author{Nobuyuki Matubayasi}
\email{nobuyuki@cheng.es.osaka-u.ac.jp}
\affiliation{Division of Chemical Engineering, Department of Materials Engineering Science, Graduate School of Engineering Science, Osaka University, Toyonaka, Osaka 560-8531, Japan}

%%%%%%%%%%%%%%%%%%%%%%%%%%%%%%%%%%%%%%%%%%%%%%%%%%%%%%%%%%%%%%%%%%%%%%%%%%%%%%%%%%%%%%%%%%%%%%%%%%%%%%%%%%%%%%%%%%%%%
\date{\today}

%%%%%%%%%%%%%%%%%%%%%%%%%%%%%%%%%%%%%%%%%%%%%%%%%%%%%%%%%%%%%%%%%%%%%%%%%%%%%%%%%%%%%%%%%%%%%%%%%%%%%%%%%%%%%%%%%%%%%

\begin{abstract}
A method for obtaining appropriate reaction coordinates is required to
identify transition states distinguishing product and reactant in complex molecular systems.
Recently, abundant research has been devoted to obtaining reaction coordinates
using artificial neural networks from deep learning literature,
where many collective variables are typically utilized in the input layer.
However, it is difficult to explain the details of which collective variables
contribute to the predicted reaction coordinates owing to the complexity
of the nonlinear
functions in deep neural networks.
To overcome this limitation,
we used Explainable Artificial
Intelligence (XAI) methods of the Local Interpretable
Model-agnostic Explanation (LIME) and
the game theory-based framework known as Shapley Additive exPlanations (SHAP).
We demonstrated that XAI enables us to obtain the degree of contribution
of each collective variable to reaction coordinates that is determined by
nonlinear regressions with deep learning for the committor of the alanine
dipeptide isomerization in vacuum.
In particular, both LIME and SHAP provide important features
to the predicted reaction
coordinates, which are characterized by appropriate dihedral angles consistent
with those previously reported from the committor test analysis.
The present study offers an AI-aided framework to explain the appropriate reaction
coordinates, which acquires considerable significance when the number of
degrees of freedom increases.
\end{abstract}

%%%%%%%%%%%%%%%%%%%%%%%%%%%%%%%%%%%%%%%%%%%%%%%%%%%%%%%%%%%%%%%%%%%%%%%%%%%%%%%%%%%%%%%%%%%%%%%%%%%%%%%%%%%%%%%%%%%%%
\maketitle

%%%%%%%%%%%%%%%%%%%%%%%%%%%%%%%%%%%%%%%%%%%%%%%%%%%%%%%%%%%%%%%%%%%%%%%%%%%%%%%%%%%%%%%%%%%%%%%%%%%%%%%%%%%%%%%%%%%%%
\section{Introduction}
\label{sec:introduction}

Identifying reaction coordinates (RCs) from a large number
of collective variables (CVs) is important for appropriately describing 
the transition state (TS) distinguishing reactant and product in various
complex molecular systems.~\cite{peters2017Reaction, krivov2013Reaction,
li2014Recent, peters2013Reaction,
wales2015Perspective, peters2016Reaction, banushkina2016Optimal,
pietrucci2017Strategies, bolhuis2021Transition}
The analysis of committor $p_\mathrm{B}(\bm{r})\in [0, 1]$, that is,
the probability of trajectories reaching the product B prior to the
reactant A starting
from any conformation $\bm{r}$, is a promising statistical examination
to explore proper 
RCs from transition path samplings using molecular dynamics (MD)
simulations.~\cite{bolhuis2002Transition, rogal2021Reaction}
The distribution of $p_\mathrm{B}$ should be unimodal with a sharp peak
at $p_\mathrm{B} \sim 0.5$ corresponding to TS because 
an appropriate RC is of such $p_\mathrm{B}$ near TS.~\cite{du1998Transition, geissler1999Kinetic,
bolhuis2000Reaction, hagan2003Atomistic, hummer2004Transition,
pan2004Dynamics, rhee2005OneDimensional, berezhkovskii2005Onedimensional,
best2005Reaction, moroni2005Interplay, peters2006Using, branduardi2007Free,
quaytman2007Reaction, antoniou2009Stochastic, peters2010TP, ernst2017Identification}

Peters \textit{et al.} have proposed 
the likelihood maximization method 
to find RCs using transition path sampling in MD
simulations.~\cite{peters2006Obtaining, peters2007Extensions, peters2010Recent}
Specifically, 
one in which the committor is modeled as a sigmoidal function,
$p_\mathrm{B}(q)=(1+\tanh(q))/2$, where $q$ is given by linear
combinations of CVs.
A good RC is obtained such that the logarithmic form of the likelihood
is maximized, corresponding to logistic regression.
This likelihood maximization method has been applied to various complex
systems.~\cite{beckham2007SurfaceMediated, beckham2008Evidence,
peters2010TransitionState, 
vreede2010Predicting, lechner2010Nonlinear, pan2010Molecular,
beckham2011Optimizing, peters2012Inertial, xi2013Hopping,
jungblut2013Optimising, mullen2014Transmission, mullen2015Easy,
lupi2016Preordering, jung2017Transition, joswiak2018Ion,
diazleines2018Maximum, okazaki2019Mechanism, arjun2019Unbiased, liang2020Identification,
rogers2020Breakage, schwierz2020Kinetic,
levintov2021Reaction, silveira2021Transitiona}
The cross-entropy minimization method has recently been proposed with
the help of the pre-evaluated committor values $p_\mathrm{B}^*$ ranging
from 0 (toward state A) to
1 (toward state B).~\cite{mori2020Dissecting, mori2020Learning}
This is an extension of the likelihood maximization in the sense that
the cross-entropy is derived from the Kullback--Leibler divergence by
considering the logarithmic form of the likelihood.~\cite{mori2020Dissecting}
In general, these maximization or minimization methods can be categorized as the linear regression
(LR) in the field of machine learning.

Machine learning models have been widely utilized to
determine the dominant CVs
from trajectories obtained using MD
simulations.~\cite{ma2005Automatic, sultan2018Automated, wehmeyer2018Timelagged, mardt2018VAMPnets,
bittracher2018Datadriven, chen2018Molecular, ribeiro2018Reweighted, 
rogal2019NeuralNetworkBased, bonati2020DataDriven, wang2020Machine,
wang2021State, sidky2020Machine,
zhang2021Deep, frassek2021Extended, hooft2021Discovering, bonati2021Deepa,
chen2021Collective, belkacemi2022Chasing}
Furthermore, a feasible application is the nonlinear regression based
on a deep neural network (DNN), which is expected to have a performance 
beyond that of the LR in searching for an appropriate
RC.~\cite{jung2019Artificial, frassek2021Extended, jung2021Autonomous, neumann2022Artificial}
In particular, nonlinear functions of a DNN with hidden layers will
provide richer expressions when the number of CVs is
drastically increased in the system of interest.
However, it remains difficult to obtain a human-interpretable
explanation for DNN learning.
Frassek \textit{et al.} reported the application of an autoencoder
consisting of an encoder, a reconstruction decoder, and a committor decoder 
to obtain a low-dimensional representation of RCs discovered from many
input CVs.~\cite{frassek2021Extended}
Jung \textit{et al.} proposed an advanced sampling scheme for rare
events, in which the maximum likelihood method combined with deep learning was designed to
identify the relevant RC.
Symbolic regression was further utilized 
to provide human-interpretable forms for trained DNN models using 
mathematical expressions.~\cite{jung2019Artificial, jung2021Autonomous}
More recently, Neumann and Schwierz applied a related model of
DNN to predict RCs of magnesium binding to RNA and used the permutation
importance method to characterize the feature importances out of
input CVs.~\cite{neumann2022Artificial}

%%%%%%%%%%%%%%%%%%%%%%%%%%%%%%%%%%%%%%%%%%%%%%%%%%%%%%%%%%%%%%%%%%%%%%%%%%%%%%%%%%%%%%%%%%
%%%%%%%%% Fig. 1 %%%%%%%%%%%%%%%%%%%%%%%%%%%%%%%%%%%%%%%%%%%%%%%%%%%%%%%%%%%%%%%%%%%%%%%%%
\begin{figure*}[t]
\includegraphics[width=0.9\textwidth]{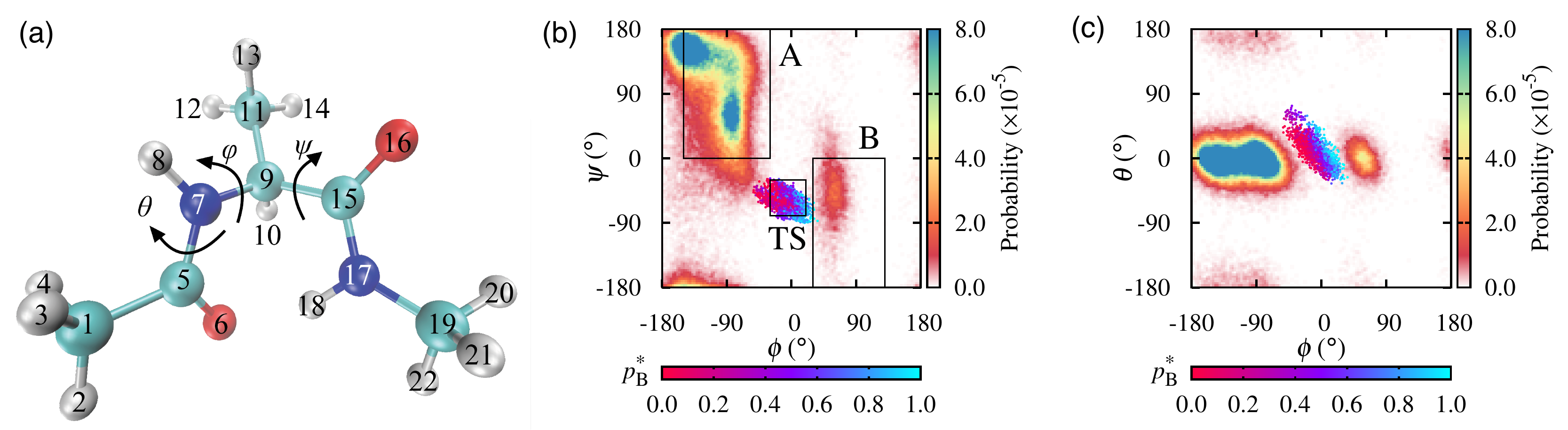}
\caption{
(a) Index assigned to alanine dipeptide atom.
Major three dihedral angles $\varphi$, $\psi$, and $\theta$ are also described.
(b) 
Ramachandran plot describing the probability distribution using 
 $\varphi$ and $\psi$.
The black boxes describe states A [$( -150^\circ,0^\circ ) \le ( \varphi,\psi ) \le (
30^\circ,180^\circ )$], B [$( 30^\circ,-180^\circ ) \le ( \varphi,\psi ) \le (
130^\circ,0^\circ )$], and TS [$(
-30^\circ,-80^\circ )\le ( \varphi,\psi ) \le ( 20^\circ,-30^\circ )$].
(c) Contour plot of the probability distribution using $\varphi$ and $\theta$.
In (b) and (c), 
the points are colored by the $p_\mathrm{B}^*$ values given in the bottom color bar.
}
\label{fig:map}
\end{figure*}
%%%%%%%%%%%%%%%%%%%%%%%%%%%%%%%%%%%%%%%%%%%%%%%%%%%%%%%%%%%%%%%%%%%%%%%%%%%%%%%%%%%%%%%%%%

%%%%%%%%%%%%%%%%%%%%%%%%%%%%%%%%%%%%%%%%%%%%%%%%%%%%%%%%%%%%%%%%%%%%%%%%%%%%%%%%%%%%%%%%%%
%%%%%%%%% Table 1 %%%%%%%%%%%%%%%%%%%%%%%%%%%%%%%%%%%%%%%%%%%%%%%%%%%%%%%%%%%%%%%%%%%%%%%%%
\begin{table*}[t]
\caption{Definition of the CV index.
The atom index is represented in Fig.~\ref{fig:map}(a). 
Note that the dihedral angles are used in cosine
 and sine forms, \textit{i.e.}, $x_{1}$ to $x_{45}$ and
 $x_{46}$ to $x_{90}$ are the cosine and sine forms,
 respectively.}
\centering
      \begin{tabular}{c|c|c|c}

      \hline
      index of CVs & \multicolumn{3}{c}{index of atoms for dihedral angles} \\ \hline
      $1-3$ &  2 -  1 -  5 -  6 &  2 -  1 -  5 -  7  &  3 -  1 -  5 -  6 \\
      $4-6$ &  3 -  1 -  5 -  7 &  4 -  1 -  5 -  6  &  4 -  1 -  5 -  7 \\
      $7-9$ &  1 -  5 -  7 -  8 &  1 -  5 -  7 -  9  &  6 -  5 -  7 -  8 \\
      $10-12$ &  6 -  5 -  7 -  9 &  5 -  7 -  9 - 10  &  5 -  7 -  9 - 11 \\
      $13-15$ &  5 -  7 -  9 - 15 &  8 -  7 -  9 - 10  &  8 -  7 -  9 - 11 \\
      $16-18$ &  8 -  7 -  9 - 15 &  7 -  9 - 11 - 12  &  7 -  9 - 11 - 13 \\
      $19-21$ &  7 -  9 - 11 - 14 & 10 -  9 - 11 - 12  & 10 -  9 - 11 - 13 \\
      $22-24$ & 10 -  9 - 11 - 14 & 15 -  9 - 11 - 12  & 15 -  9 - 11 - 13 \\
      $25-27$ & 15 -  9 - 11 - 14 &  7 -  9 - 15 - 16  &  7 -  9 - 15 - 17 \\
      $28-30$ & 10 -  9 - 15 - 16 & 10 -  9 - 15 - 17  & 11 -  9 - 15 - 16 \\
      $31-33$ & 11 -  9 - 15 - 17 &  9 - 15 - 17 - 18  &  9 - 15 - 17 - 19 \\
      $34-36$ & 16 - 15 - 17 - 18 & 16 - 15 - 17 - 19  & 15 - 17 - 19 - 20 \\
      $37-39$ & 15 - 17 - 19 - 21 & 15 - 17 - 19 - 22  & 18 - 17 - 19 - 20 \\
      $40-42$ & 18 - 17 - 19 - 21 & 18 - 17 - 19 - 22  &  1 -  7 -  5 -  6 \\
      $43-45$ &  5 -  9 -  7 -  8 &  9 - 17 - 15 - 16  & 15 - 19 - 17 - 18 \\ \hline

      \end{tabular}
      \label{tab:dihed} 
\end{table*}
%%%%%%%%%%%%%%%%%%%%%%%%%%%%%%%%%%%%%%%%%%%%%%%%%%%%%%%%%%%%%%%%%%%%%%%%%%%%%%%%%%%%%%%%%%

In this study, we propose an artificial intelligence (AI)-aided method
to determine the nonlinear RC and then interpret the
RC locally at the TS using an explainable AI (XAI) framework.
DNN is used to identify the appropriate
RC from the CV dataset and pre-evaluated committor values
$p_\mathrm{B}^*$ obtained from transition path samplings.
The target reaction is the C$_\mathrm{7eq}$ and C$_\mathrm{7ax}$ isomerization of 
alanine dipeptide in vacuum (see Fig.~\ref{fig:map}(a)).
Empirically, the dihedral angle change in $\varphi$ is thought to be coupled with
the other major angle $\psi$.
Bolhuis \textit{et al.} revealed that an additional
dihedral angle $\theta$ next to $\varphi$ becomes relevant for describing the
proper committor $p_\mathrm{B}$ distribution with a peak at
$p_\mathrm{B}\sim 0.5$.~\cite{bolhuis2000Reaction}
An analogous result was reported by Ren \textit{et al.} using the string method.~\cite{ren2005Transition}
Furthermore, in a seminal study by Ma and Dinner,
a neural network combined with a genetic algorithm was applied to the committor values, 
predicting that an appropriate RC involves the dihedral angle
$\theta$.~\cite{ma2005Automatic}
We also demonstrated that 
the cross-entropy minimization enabled elucidation of the importance of
$\theta$ in the alanine dipeptide isomerization.~\cite{mori2020Learning}
More recently, Manuchehrfar \textit{et al.} reported persistent homology
analysis results for characterizing configurations with committor values
$p_\mathrm{B} \sim 0.5$ on a two-dimensional plot of probability density using
$\varphi$ and $\theta$.~\cite{manuchehrfar2021Exacta}
Note that there are other target reactions that can be applied
by the DNN, such
as the Rate-Promoting Vibrations Model of enzyme
catalysis,~\cite{antoniou2004Transition} of which RC was identified using
the likelihood maximization
by Peters.~\cite{peters2010TransitionState}

The purpose of XAI is 
to provide an explainable model for the black-box-type predictions of
DNNs.~\cite{adadi2018Peeking, molnar2020Interpretable}
In other words, XAI can be regarded as a class of 
model-agnostic interpretation method, which is separated from predictions using DNN.
XAI is further classified into local and global explanation methods. 
Global explanation methods, including the permutation
importance method used in Ref.~\onlinecite{neumann2022Artificial},
characterize the average contribution of input
variables to the prediction.
In contrast, local explanation methods have an advantage of giving interpretable
models to the individual predictions by 
estimating the contribution of each input variable to each prediction.
As major implementations of local explanation methods, the Local Interpretable
Model-agnostic Explanation (LIME)~\cite{ribeiro2016Whya} and 
the game theory-based framework known as Shapley Additive exPlanations
(SHAP)~\cite{lundberg2017unified} are employed.
We locally examined the feature contributions of the input CVs to an
appropriate RC predicted from the DNN using LIME and SHAP.
In particular, it is important to obtain a local explanation model 
for conformations that exhibit $p_\mathrm{B}^* \sim 0.5$ corresponding to TS.
This information may not be readily accessed by the symbolic regression,
which provides a global model for RC.
Finally, the extracted dominant CVs are interpreted by a separatrix line
distinguishing states A and B on the contour map of the 
probability distribution using two dihedral angles, $\varphi$ and $\theta$.

\section{Methods}
\label{sec:method}

%%%%%%%%%%%%%%%%%%%%%%%%%%%%%%%%%%%%%%%%%%%%%%%%%%%%%%%%%%%%%%%%%%%%%%%%%%%%%%%%%%%%%%%%%%
%%%%%%%%% Fig. 2 %%%%%%%%%%%%%%%%%%%%%%%%%%%%%%%%%%%%%%%%%%%%%%%%%%%%%%%%%%%%%%%%%%%%%%%%%
\begin{figure*}[t]
\centering
\includegraphics[width=0.65\textwidth]{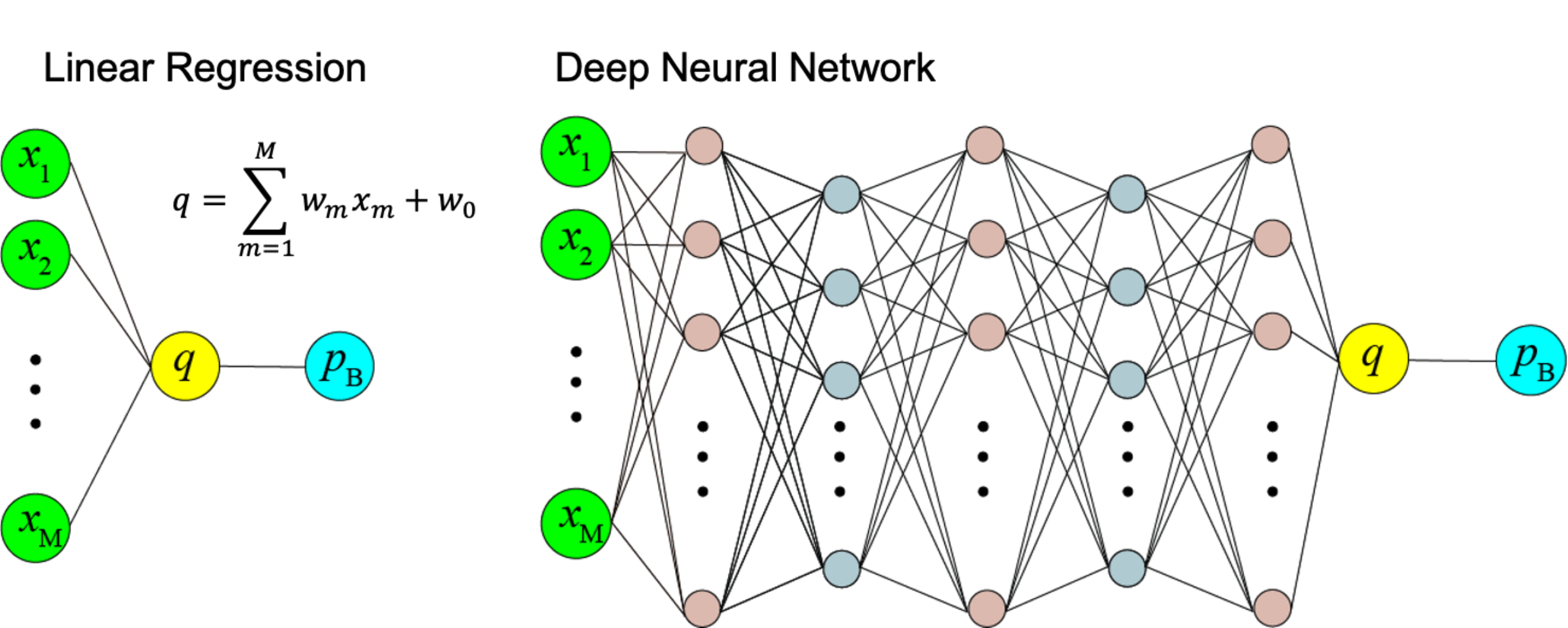}
\caption{
Schematic diagram of LR and DNN learning the relationship between
 pre-evaluated 
committor value
 $p_\mathrm{B}^*$ and $M=90$ CVs plus 1 bias term $\bm{x}=(1, x_1, x_2, \cdots, x_M)$, and predicting committor
 $p_\mathrm{B}$ as a sigmoidal function
$p_\mathrm{B}(q)=(1+\tanh(q))/2$. 
Note that the node representing the bias term is omitted from the diagram.
}
\label{fig:dnn}
\end{figure*}
%%%%%%%%%%%%%%%%%%%%%%%%%%%%%%%%%%%%%%%%%%%%%%%%%%%%%%%%%%%%%%%%%%%%%%%%%%%%%%%%%%%%%%%%%%

\subsection{Simulation details}

We numerically examined the isomerization of alanine dipeptide in a
vacuum using MD simulations. 
The system is the same as that in our previous study, where
one alanine dipeptide molecule was simulated (other numerical conditions are
described in
Ref.~\onlinecite{mori2020Learning}). 

Figure~\ref{fig:map}(b) shows the Ramachandran plot of the alanine
dipeptide using the major dihedral angles, $\varphi$(C-N-C$_\alpha$-C) and
$\psi$(N-C$_\alpha$-C-N) (see also Fig.~\ref{fig:map}(a)).
We examined the transition paths between 
two energetically stable states, the $\beta$-sheet structure
(C$_\mathrm{7eq}$ denoted as state A) and the left-handed $\alpha$-helix
structure (C$_\mathrm{7ax}$ denoted as state B), which are
characterized in Fig.~\ref{fig:map}(b).
We defined states A and B as 
A [$( -150^\circ,0^\circ ) \le ( \varphi,\psi ) \le (
30^\circ,180^\circ )$] and 
B [$( 30^\circ,-180^\circ ) \le ( \varphi,\psi ) \le (
130^\circ,0^\circ )$], respectively.
In addition, the intermediate region was regarded as TS [$(
-30^\circ,-80^\circ )\le ( \varphi,\psi ) \le ( 20^\circ,-30^\circ )$].

In our previous study,~\cite{mori2020Learning} we sampled 2,000 shooting
points from the TS region using the
aimless shooting method.~\cite{peters2006Obtaining} 
For each shooting point, the velocity was randomly assigned according to
the Maxwell--Boltzmann distribution at 300 K, generating a trajectory of
1 ps.
This was repeated 100 times for each point, and the committor value $p_\mathrm{B}^*$
was quantified from the number of transitions to states A or B.
We also calculated a total of 90 CVs from all 45 dihedral angles in the
molecule into cosine and sine forms.
See Fig.~\ref{fig:map}(a) and Table~\ref{tab:dihed} for details of the
investigated dihedral angles.
We used the same datasets of CVs and $p_\mathrm{B}^*$ as those used in
Ref.~\onlinecite{mori2020Learning}.

The shooting points with $0.45 \le p_\mathrm{B}^* \le 0.55$ are shown in
Fig.~\ref{fig:map}(b) and (c).
Figure~\ref{fig:map}(c) shows the contour map of the
probability distribution as a function of $\varphi$ and $\theta$.
Although points with $p_\mathrm{B}^* \sim 0.5$ are widely distributed on the
$(\varphi,\psi)$ plane, a clear separatrix line can be described on the
$(\varphi,\theta)$ plane.
This result indicates that the appropriate RC is related to $\theta$ 
instead of $\psi$.~\cite{mori2020Learning}

\subsection{Linear regression and deep neural network}

In this study, we used the LR and 
DNN to learn the relationship between the committor distribution $p_\mathrm{B}^*$
and candidate CVs, yielding the committor prediction.
Figure~\ref{fig:dnn} illustrates a schematic of the training of the LR and DNN.
The procedure consists of two parts:
one is the training part that
transforms from the input layer given by the CVs into a one-dimensional variable $q$,
and the other is the prediction part that transforms $q$ into the
sigmoidal function
$p_\mathrm{B}(q)=(1+\tanh(q))/2$.

The training part for both LR and DNN is set up as follows:
LR is implemented by a simple perceptron, and thus the output can be
described by $q = \sum_{m=1}^{M} w_m x_m+w_0$.
Here, $M$ is the number of CVs, $x_m$ is the $m$-th CV, and $w_m$ is the
corresponding coefficient.
In addition, $w_0$ denotes the bias term.
By contrast, 
the DNN consisted of five hidden layers, of which the odd- and even-numbered layers had 400
and 200 nodes, respectively.
We used the leaky rectified linear unit (Leaky ReLU) with a leaky parameter
set to 0.01 as the default for the activation function.~\cite{maas2013Rectifier}
The output is a one-dimensional variable $q$.
For the prediction part for both the LR and DNN, 
the relationship between the output $q$ and committor $p_\mathrm{B}$ is
given by $p_\mathrm{B}(q)=(1+\tanh(q))/2$.
In other words, the corresponding activation function can be described in a sigmoidal
manner.

The dataset of CVs and committor values from 2,000 shooting points were divided into
training, validation, and test datasets at a ratio of 5:1:4.
The variables of CVs were standardized.
Optimization was performed using AdaMax.~\cite{kingma2017Adam}
The learning rate $lr$ and two decay factors $\beta_1$ and $\beta_2$
were set to the default values of 0.001, 0.9, and 0.99, respectively.
The $L_2$ norm regularization was set to both LR and DNN with a
regularization parameter of $0.001$ to prevent overfitting.
In addition, the dropout was set to the hidden layers at a rate of
0.5 during DNN training.
We used the TensorFlow library to implement the LR and DNN.~\cite{abadi2016TensorFlow}

The cross-entropy function~\cite{mori2020Dissecting, mori2020Learning}
\begin{align}
&\mathcal{H}\left(p_\mathrm{B}^*, p_\mathrm{B}\right)\nonumber\\
&\quad = -\sum_{k=1}^{N} p_\mathrm{B}^*\left(\mathbf{r}_k\right)\ln{p_\mathrm{B}(q)}
 - \sum_{k=1}^{N}
 \left(1-p_\mathrm{B}^*\left(\mathbf{r}_k\right)\right)\ln{\left[1-p_\mathrm{B}(q)\right]}, 
\label{eq:cross_entropy}
\end{align}
was used to derive the loss function for both the LR and DNN training.
Here, $\bm{r}_k$ represents the $k$-th conformation of the molecule with the number of
shooting points $N$.
The cross-entropy minimization principle was derived from the Kullback--Leibler
divergence~\cite{mori2020Dissecting} and was applied to 
the LR to search for the appropriate RC for alanine dipeptide
isomerization.~\cite{mori2020Learning}

%%%%%%%%%%%%%%%%%%%%%%%%%%%%%%%%%%%%%%%%%%%%%%%%%%%%%%%%%%%%%%%%%%%%%%%%%%%%%%%%%%%%%%%%%%
%%%%%%%%% Fig. 3 %%%%%%%%%%%%%%%%%%%%%%%%%%%%%%%%%%%%%%%%%%%%%%%%%%%%%%%%%%%%%%%%%%%%%%%%%
\begin{figure}[t]
\centering
\includegraphics[width=0.45\textwidth]{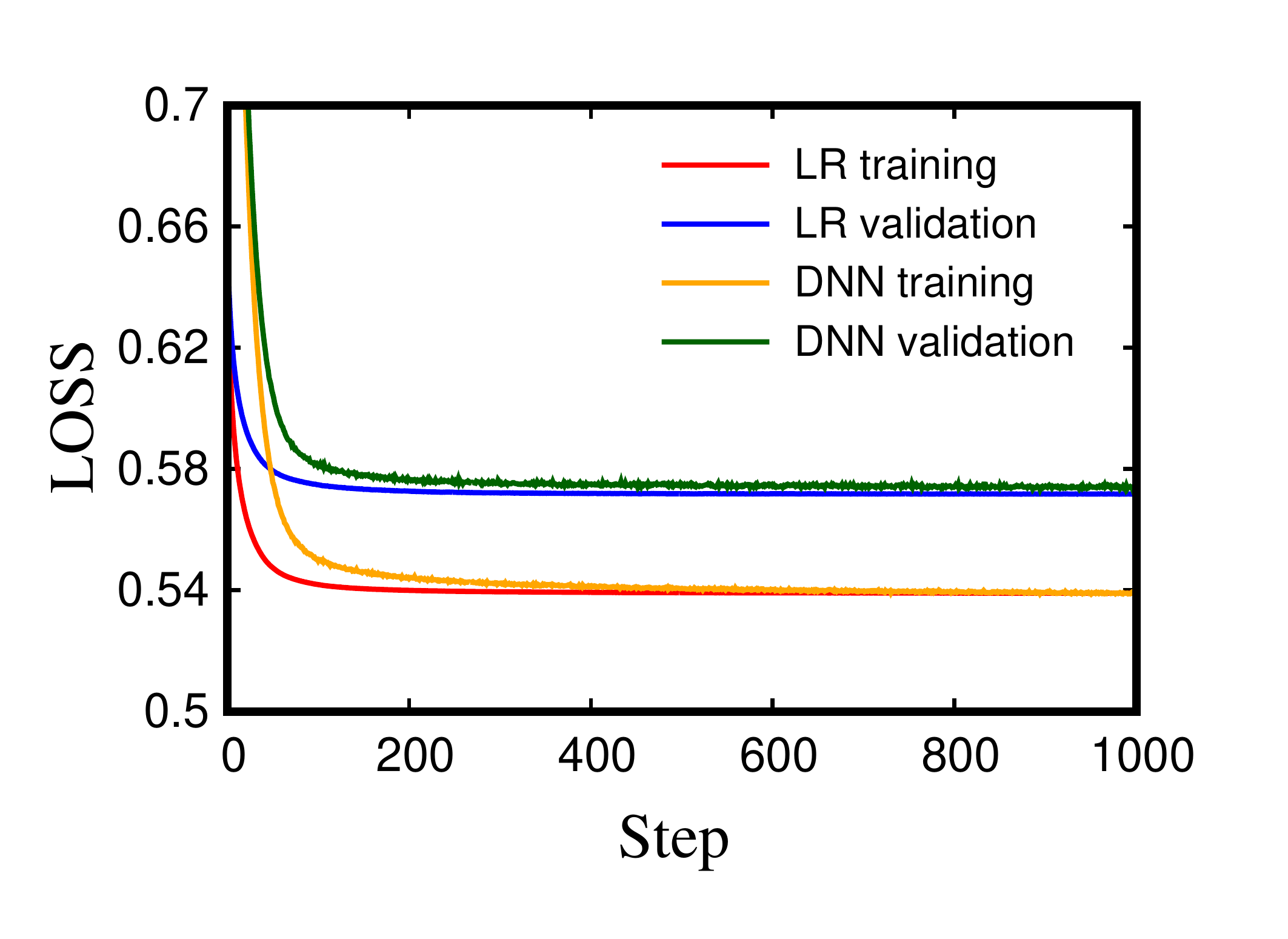}
\caption{
Training process of cross-entropy minimization for both LR and DNN. 
Red and blue (orange and green) curves indicate the loss function 
 (cross-entropy $\mathcal{H}$ in Eq.~(\ref{eq:cross_entropy})) of LR (DNN) as a function of iteration
 steps with training and validation dataset,
 respectively.
}
\label{fig:loss} 
\end{figure}
%%%%%%%%%%%%%%%%%%%%%%%%%%%%%%%%%%%%%%%%%%%%%%%%%%%%%%%%%%%%%%%%%%%%%%%%%%%%%%%%%%%%%%%%%%

%%%%%%%%%%%%%%%%%%%%%%%%%%%%%%%%%%%%%%%%%%%%%%%%%%%%%%%%%%%%%%%%%%%%%%%%%%%%%%%%%%%%%%%%%%
%%%%%%%%% Fig. 4 %%%%%%%%%%%%%%%%%%%%%%%%%%%%%%%%%%%%%%%%%%%%%%%%%%%%%%%%%%%%%%%%%%%%%%%%%
\begin{figure}[t]
\centering
\includegraphics[width=0.45\textwidth]{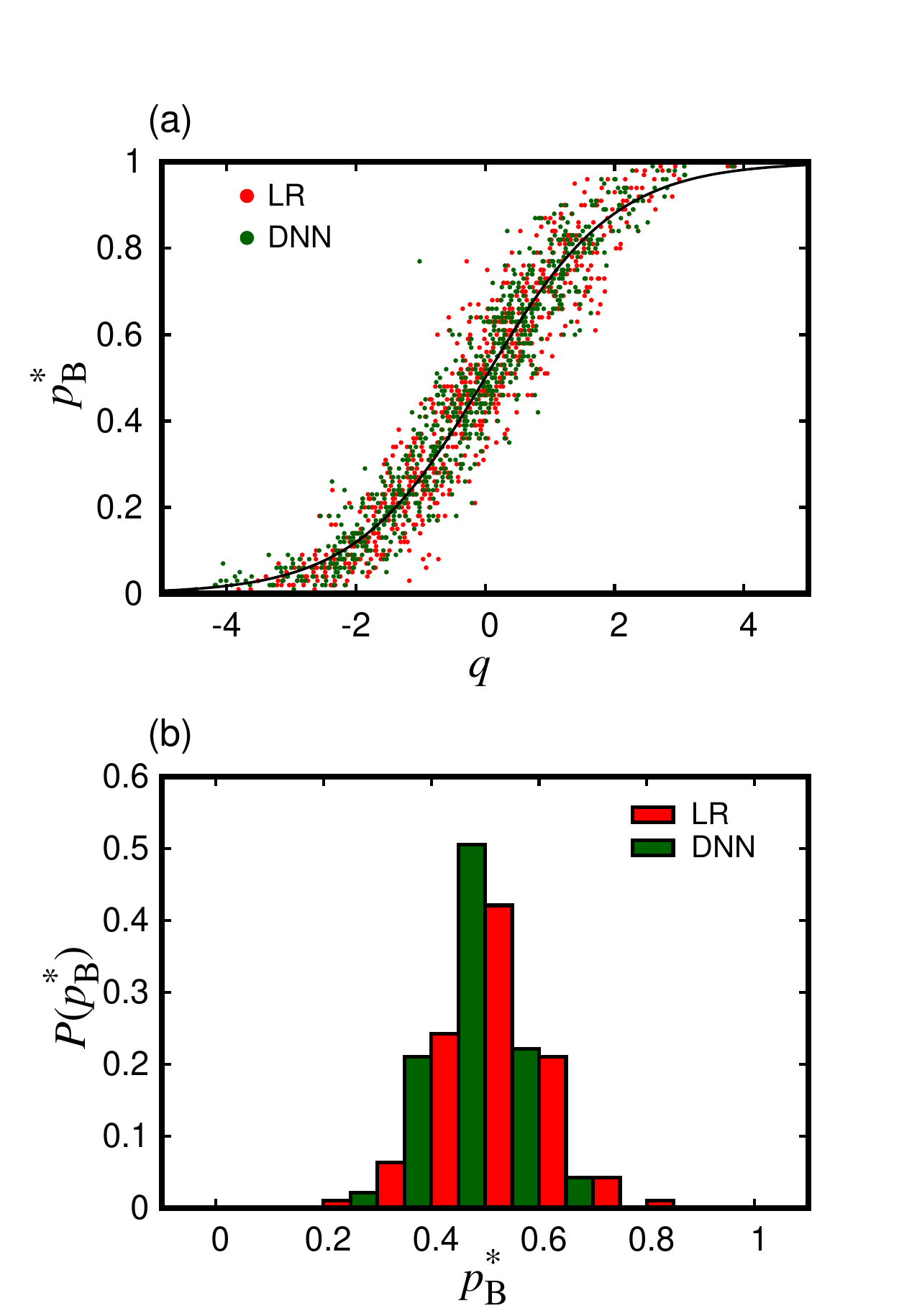}
\caption{
(a) Relationship between committor distributions $p_\mathrm{B}^*$ and
 $q$ obtained from LR (red) and DNN (green) trained model using test
 dataset (800 points). 
The solid black line represents the sigmoidal function, $p_\mathrm{B}(q)=(1+\tanh(q))/2$. 
(b) Probability of $p_\mathrm{B}^*$ for $q\in (-0.2, 0.2)$ for LR (red)
 and DNN (green), where the
 points are extracted from the data in (a).
}
\label{fig:prd} 
\end{figure}
%%%%%%%%%%%%%%%%%%%%%%%%%%%%%%%%%%%%%%%%%%%%%%%%%%%%%%%%%%%%%%%%%%%%%%%%%%%%%%%%%%%%%%%%%%

\subsection{LIME - Local Interpretable Model-agnostic Explanation} 

LIME was applied to explain the committor predictions obtained by the DNN training.
In general, DNNs are very complex in terms of obtaining understandable features.
LIME can explain the contribution of each input variable
of any black-box-type classifier using a more interpretable model.~\cite{ribeiro2016Whya}
It provides a
linear regression function for
the local behavior of a target instance explained by perturbation of input variables.
In other words, features with large coefficients in the linear
regression function provide a predictive interpretation.
The interpretable feature $\xi(\bm{x})$ is obtained by a linear
regression function $g$ for the input
data by the CV vector $\bm{x}=(1, x_1, x_2, \cdots, x_M)$ (90
dihedral angles plus 1 bias term in our case) and the black-box model
$f$ by the DNN.
This is represented by the following equation,
\begin{align}
\xi(\bm{x}) =  \argmin_{g \in G} \left( L^\mathrm{LIME}(f, g, \pi_{\bm{x}}^\mathrm{LIME})  + \Omega(g)\right),
\label{eq:LIME}
\end{align}
where $L^\mathrm{LIME}(f, g,\pi_{\bm{x}}^\mathrm{LIME})$ is the squared loss function that measures the distance
between $f$ and $g$ with 
$G$ representing a class of explanation models ($g\in G$).
$\pi_{\bm{x}}^\mathrm{LIME}$ represents the proximity measure around the input data $\bm{x}$
to be explained.
In practice, the exponential kernel function is
\begin{align}
\pi_{\bm{x}}^\mathrm{LIME}(\bm{z}) = \exp{\left( -D(\bm{x},\bm{z})^2/\sigma^2
\right)},
\label{eq:LIME_weight}
\end{align}
where the distance function $D$ and width $\sigma$
are used for any perturbed instance $\bm{z}$, which is randomly
generated around the data $\bm{x}$.
In LIME,
the perturbed data $\bm{z}$ around $\bm{x}$ weighted by the proximity
measure is transformed into binary
variables $\bm{z}' \in \{0, 1\}^M$ with the number of input variables
$M$ for the human-understandable presentation because the important components of the original input data are
not always interpretable.
Regularization $\Omega(g)$ is also used 
to reduce the complexity of the explanation for $g \in G$.
We used the LIME package, which is available at \url{https://github.com/marcotcr/lime}.

%%%%%%%%%%%%%%%%%%%%%%%%%%%%%%%%%%%%%%%%%%%%%%%%%%%%%%%%%%%%%%%%%%%%%%%%%%%%%%%%%%%%%%%%%%
%%%%%%%%% Fig. 5 %%%%%%%%%%%%%%%%%%%%%%%%%%%%%%%%%%%%%%%%%%%%%%%%%%%%%%%%%%%%%%%%%%%%%%%%%
\begin{figure*}[t]
\centering
\includegraphics[width=0.9\textwidth]{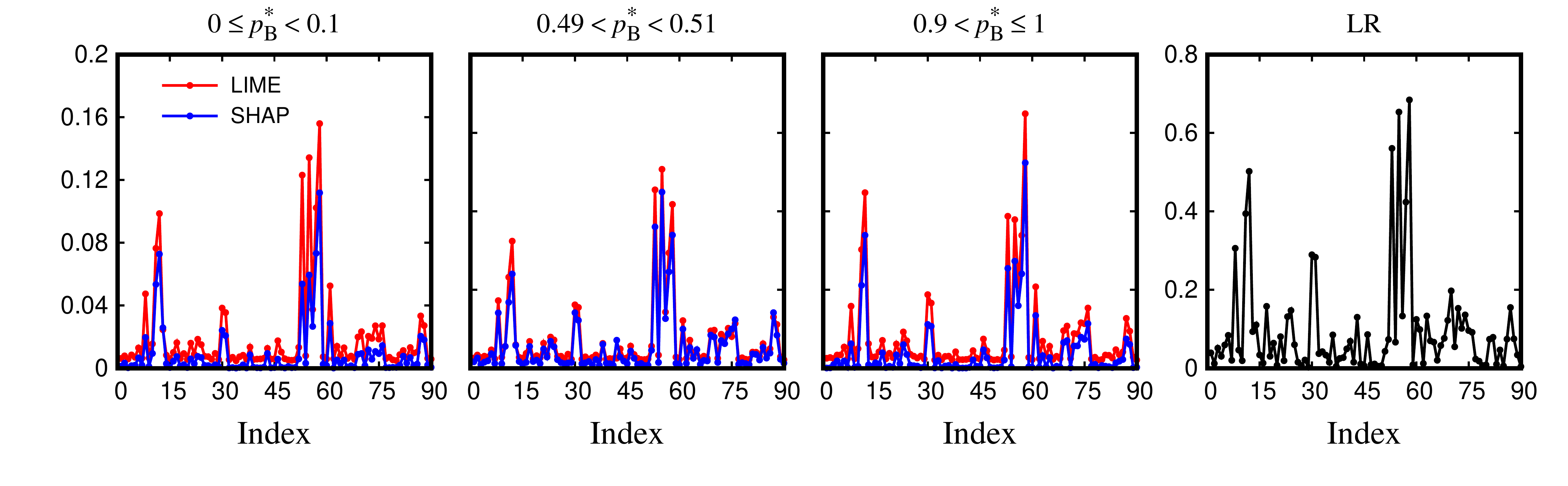}
\caption{
Feature contribution of each collective variable in absolute value
 obtained using LIME (red) and SHAP (blue) for the DNN training.
The local explanation is provided for three regions corresponding to $0\le p_\mathrm{B}^*< 0.1$, $0.49< p_\mathrm{B}^*
 <0.51$, and $0.9 < p_\mathrm{B}^*\le 1$ from left to right.
The rightmost panel shows optimized coefficients $w_m$ of each
 variable $x_m$ in absolute value obtained from the LR model.
}
\label{fig:localparam} 
\end{figure*}
%%%%%%%%%%%%%%%%%%%%%%%%%%%%%%%%%%%%%%%%%%%%%%%%%%%%%%%%%%%%%%%%%%%%%%%%%%%%%%%%%%%%%%%%%%

%%%%%%%%%%%%%%%%%%%%%%%%%%%%%%%%%%%%%%%%%%%%%%%%%%%%%%%%%%%%%%%%%%%%%%%%%%%%%%%%%%%%%%%%%%
%%%%%%%%% Table 2 %%%%%%%%%%%%%%%%%%%%%%%%%%%%%%%%%%%%%%%%%%%%%%%%%%%%%%%%%%%%%%%%%%%%%%%%%
\begin{table*}[t]
\caption{Top five dominant contributions and those absolute values obtained
 using LIME and SHAP for the three regions, $0\le p_\mathrm{B}^*< 0.1$, $0.49< p_\mathrm{B}^*
 <0.51$, and $0.9 < p_\mathrm{B}^*\le 1$.
The result of the LR model is also shown for comparison.
}
\centering
      \begin{tabular}{c|c|c|c|c|c|c|c|c|c|c|c|c|c}
     \hline 
	\multicolumn{4}{c}{ $0\le p_\mathrm{B}^* < 0.1$}&\multicolumn{4}{|c|}{
       $0.49< p_\mathrm{B}^* <0.51$}&\multicolumn{4}{|c|}{ $0.9 <
       p_\mathrm{B}^* \le 1$}&\multicolumn{2}{c}{ $0 \le p_\mathrm{B}^* \le 1$}\\
      \hline
\multicolumn{2}{c|}{LIME} &\multicolumn{2}{c|}{SHAP}&\multicolumn{2}{c|}{LIME} &\multicolumn{2}{c|}{SHAP}&\multicolumn{2}{c|}{LIME} &
\multicolumn{2}{c|}{SHAP}&\multicolumn{2}{|c}{LR}\\ \hline
  index & value& index & value& index & value& index & value&index & value&index & value&index&value\\ \hline
57&0.156&57&0.112&54&0.127&54&0.112&57&0.162&57&0.131&57&0.684\\ 
54&0.134&56&0.073&52&0.114&52&0.090&11&0.112&11&0.085&54&0.653\\ 
52&0.123&11&0.073&57&0.104&57&0.085&52&0.097&54&0.068&52&0.561\\ 
56&0.102&54&0.059&11&0.081&56&0.061&54&0.095&52&0.064&11&0.502\\ 	
11&0.099&52&0.054&56&0.074&11&0.060&56&0.085&56&0.060&56&0.424\\ 

      \hline

      \end{tabular}
      \label{tab:xai_p} 
\end{table*}
%%%%%%%%%%%%%%%%%%%%%%%%%%%%%%%%%%%%%%%%%%%%%%%%%%%%%%%%%%%%%%%%%%%%%%%%%%%%%%%%%%%%%%%%%%

\subsection{SHAP - Shapley Additive exPlanations}

We also applied SHAP, which is a game-theory-based method for explaining
the predictions of
black-box-type models.~\cite{lundberg2017unified}
Note that LIME assumes that the local behavior can be
described by the linear model, but this has no
theoretical background.
In contrast, SHAP guarantees that the prediction is fairly distributed among the
input features.
In fact, SHAP utilizes the Shapley value,~\cite{shapley195317} which is 
a value that fairly distributes the rewards given by the team
collaboration to individual players.
In SHAP, an additive feature attribution method provides a linear
function consisting of binary variables:
\begin{align}
g(\bm{z}') = \phi_0 + \sum_{i = 1}^M \phi_i z'_i,
\label{eq:additive_feature}
\end{align}
where the coefficient $\phi_i$ serves as the Shapley value, which explains
the importance of the binary feature $\bm{z}'\in \{0, 1\}^M$.
LIME can be understood as an additive feature attribution method that provides
linear models in the binary vector space in the sense that 
the explanation model $g$ of LIME can be expressed by 
Eq.~(\ref{eq:additive_feature}).

We used one of several implementations for SHAP, known as
the Kernel SHAP, which is designed as a model-agnostic estimation that 
provides a local explanation 
model using Shapley values and LIME.
For the Kernel SHAP, the following specific equations are used 
with Eq.~(\ref{eq:LIME}) of the LIME algorithm:
\begin{align}
\Omega(g) &=0,\label{eq:SHAP_penalty}\\
\pi_{\bm{x}}^\mathrm{SHAP}(\bm{z}')&=\frac{M-1}{{}_M
 C_{|\bm{z}'|}|\bm{z}'|(M-|\bm{z}'|)},\label{eq:SHAP_weight}\\
L^\mathrm{SHAP}(f, g, \pi_{\bm{x}}^\mathrm{SHAP}) & =\sum_{\bm{z}'\in
 Z}\pi_{\bm{x}}(\bm{z}')\left(f(h_{\bm{x}}^{-1}(\bm{z}'))-g(\bm{z}')\right)^2,
\end{align}
where $|\bm{z}'|$ is the number of non-zero features in $\bm{z}'$ and
$h_{\mathrm{x}}$ is a mapping function of binary variables $\bm{z}'$
into the original input data $\bm{x}$.
The local explanation model $g(\bm{z}')$ approximating
$f(h_{\bm{x}}^{-1}(\bm{z}'))$ can be obtained using a weighted
linear regression with Eq.~(\ref{eq:SHAP_weight}), which differs 
from the weight function used in LIME (see Eq.~(\ref{eq:LIME_weight})).
Note the contrast between Eqs.~(\ref{eq:LIME_weight}) and
(\ref{eq:SHAP_weight}), which are used in LIME and SHAP, respectively.
In practice,
if we use SHAP's kernel of Eq.~(\ref{eq:SHAP_weight}) as LIME's kernel,
LIME will provide values similar to SHAP values.
The SHAP package is available at \url{https://github.com/slundberg/shap}.
We obtained SHAP values using this package, where
the Akaike information criterion is used for the
regularization as default.

\section{Results and discussion}
\label{sec:results}

\subsection{Training and prediction of committor}

Figure~\ref{fig:loss} shows the learning process using a training
dataset (1,000 points) for both LR and DNN by
plotting the loss function as a function of the iteration step.
The loss function is the cross-entropy of Eq.~(\ref{eq:cross_entropy}) for both of LR and DNN. 
In parallel, to check for overfitting of the training model,
we monitored the loss function values obtained from the
validation dataset (200 points), which are also plotted in Fig.~\ref{fig:loss}.
To examine the robustness of the learning, we performed 10 trials from
randomly chosen coefficients and confirmed that there was no significant
difference in the converged value of the loss function for both LR and DNN. 
The results are presented as an average of 10 trials.
It is demonstrated that the loss function decreases without any
increase in those for the validation dataset and finally converges after several hundred
iterations for both LR and DNN.
The convergence of DNN is slower than that of LR, which can be
regarded as a result of the number of parameters, that is, 91 for
LR and approximately 360,000 for DNN.

The LR/DNN trained model provides a prediction with regard to the
relationship between 
the committor value $p_\mathrm{B}$ and RC $q$ from the test dataset (800 points).
Figure~\ref{fig:prd}(a) shows the committor distribution $p_\mathrm{B}^*$ as a function of
$q$ for the test dataset.
It was confirmed that $p_\mathrm{B}$ values of the test dataset globally
follow the sigmoidal function.
To further investigate $p_\mathrm{B}$ values close to 
the TS at $q=0$, the distribution of $p_\mathrm{B}$ in the range of
$-0.2 < q < 0.2$ is shown in Fig.~\ref{fig:prd}(b).
The $p_\mathrm{B}$ distribution exhibits a sharp peak at approximately 
0.5 for both LR and DNN trainings.
Thus, the predicted $q$ from both LR and DNN training can 
characterize the appropriate RC for the alanine dipeptide isomerization.

\subsection{Contribution of collective variables to the prediction}

We used XAI, that is, LIME and SHAP, to explain 
the contribution of the input CVs to the RC from our DNN training.
In practice, the dataset (2,000 points) is classified into three regions, 
$0 \le p_\mathrm{B} < 0.1$ (near state A), $0.49 < p_\mathrm{B} <0.51$ (near 
TS), and $0.9 < p_\mathrm{B} \le 1$ (near state B) to provide a local explanation.
For each region, randomly chosen 30 points were analyzed by LIME and
SHAP, from which the contributions of the input variables to $q$ (the
appropriate RC) were quantified.
Figure~\ref{fig:localparam} shows the absolute values of the feature
contributions averaged over 30 points for the three $p_\mathrm{B}$
regions obtained using LIME ans SHAP.
For comparison, the absolute values of the optimized coefficient $w_m$
obtained from the LR training are also plotted in Fig.~\ref{fig:localparam}.
The feature contribution values of LIME and SHAP have the same meaning as
the coefficient $w_m$ obtained
by the LR model, although the values do not match among LIME, SHAP, and LR.
The top five dominant contributions and their absolute values are listed in Table~\ref{tab:xai_p}. 

Indices 57 and 54, corresponding to $\sin \varphi$ (5-7-9-11) and
$\sin\theta$ (6-7-8-9), respectively, have large contributions using the
LR model, which is consistent with the results of our previous
study.~\cite{mori2020Learning}
This result indicates that the relevant angle to the rotation of $\varphi$
is not $\psi$, but instead 
$\theta$.
Notably, both LIME and SHAP reveal 
two dihedral angles, $\varphi$ (indices 11, 56, and 57) and
$\theta$ (indices 52 and 54), as major contributions to the
$p_\mathrm{B}$ prediction obtained from the DNN.
The results obtained by LIME and SHAP are similar 
over the three regions of $p_\mathrm{B}$.
This observation may justify LIME's linear model assumption by being
consistent with the SHAP's result guaranteed by the game theory.
However, the
order is different between TS ($0.49 < p_\mathrm{B}<
0.51$) and states near A or B ($0\le p_\mathrm{B} < 0.1$ or
$0.9 < p_\mathrm{B} \le 1$). 
This difference can be understood as follows.
Near the two stable states, A and B, 
the contribution of variables relating to $\varphi$ is larger than those 
corresponding to $\theta$ because a large increase in the angle $\varphi$
is necessary for transitions A $\to$ B and B $\to$ A.
By contrast, near TS, both LIME and SHAP explain that 
the contribution of $\theta$ (indices 52 and 54) becomes larger, as 
the change in $\theta$ has more influence than the change in $\varphi$. 
In Fig.~\ref{fig:map}(c), we show the two-dimensional probability distribution map
of $\varphi$ and $\theta$, where the distribution of shooting
points having $p_\mathrm{B}^*\sim 0.5$ are located not perpendicular
to $\varphi$, but tilted in the direction of $\theta$.
Thus, the $\theta$-angle change becomes important when crossing the 
separatrix distinguishing states A and B, whereas the $\varphi$-angle
change is globally important for the target isomerization.
Such a detailed process cannot be described by coefficients globally
optimized using the 
LR model, but is revealed by the local explanation model using DNN
with XAI.

\section{Conclusions}
\label{sec:conclusion}

In this paper, we proposed an AI-aided method, in which
DNN is used to identify the appropriate RC for alanine
dipeptide isomerization in vacuum.
We trained the DNN to predict the committor function in a sigmoidal
manner from the dataset of committor
$p_\mathrm{B}^*$ and dihedral angles in the cosine and sine
forms.
The DNN revealed the importance of $\theta$ rather than $\psi$
along the rotation about $\varphi$, 
which is consistent with various studies using the committor test
analysis.~\cite{bolhuis2000Reaction, ma2005Automatic, ren2005Transition}

Furthermore, LIME and SHAP were used as XAI tools 
to provide a local explanation model for black-box-type DNN
prediction, in contrast to the symbolic regression method for
obtaining the human-interpretable model.~\cite{jung2019Artificial, jung2021Autonomous}
It was demonstrated that LIME and SHAP enabled the explanation for 
three $p_\mathrm{B}$ regions,
$0 \le p_\mathrm{B} < 0.1$ (near state A), $0.49 < p_\mathrm{B} <0.51$ (near 
TS), and $0.9 < p_\mathrm{B} \le 1$ (near state B).
In particular, the feature contribution of $\theta$ becomes more evident near
TS than near state A or B, indicating the necessity of $\theta$-angle change for 
crossing the barrier between the two stable states A and B.
In fact, the influence of $\theta$ on $\varphi$-angle change is in accordance
with the separatrix line described by shooting points with $p_\mathrm{B}^* \sim 0.5$
on the probability distribution of $\varphi$ and $\theta$, which is
tilted in the $\theta$-axis direction.

Finally, it should be noted that 
understanding the detailed mechanism for 
complex molecular systems becomes complicated because of the large numbers of CVs,
regardless of the use of DNNs.
Important targets are solute-solvent systems such as 
alanine dipeptide in explicit water, which was examined using neural
networks to predict the committor by Ma and Dinner.~\cite{ma2005Automatic}
The DNN predicting the committor is further applied to various solute-solvent
systems,~\cite{jung2019Artificial, frassek2021Extended,
jung2021Autonomous, neumann2022Artificial} and
the current AI-aided method combined with XAI will be used as a practical
tool to provide a local explanation for an appropriate RC obtained from DNN
training, particularly near the TS of the transition connecting the stable states.

\begin{acknowledgments}
This work was supported by JSPS KAKENHI Grant Numbers:
 JP20J14619~(T.K.), JP18K05049~(T.M.),
JP18H01188~(K.K.), JP19H01812~(K.K.), JP20H05221~(K.K.),
 JP22H04542~(K.K.), JP22K03550~(K.K.), and JP19H04206~(N.M.).
This work was also partially supported 
by the Fugaku Supercomputing Project (No.~\mbox{JPMXP1020200308}) and the Elements Strategy
Initiative for Catalysts and Batteries (No.~\mbox{JPMXP0112101003}) from the
Ministry of Education, Culture, Sports, Science, and Technology.
T.M. thanks the Pan-Omics Data-Driven Research Innovation Center, Kyushu University for financial support.
The numerical calculations were performed at Research Center of
Computational Science, Okazaki Research Facilities, National Institutes
of Natural Sciences (Project: 21-IMS-C058) and at the Cybermedia Center, Osaka University.
\end{acknowledgments}

\section*{AUTHOR DECLARATIONS}

\section*{Conflicts of Interest}
The authors have no conflicts to disclose.

\section*{Data availability statement}

The data that support the findings of this study are openly available in
Zenode at \url{https://doi.org/10.5281/zenodo.6392326}.

%aipnum4-2.bst 2019-01-14 (MD) hand-edited version of apsrev4-1.bst


\begin{thebibliography}{87}%
\makeatletter
\providecommand \@ifxundefined [1]{%
 \@ifx{#1\undefined}
}%
\providecommand \@ifnum [1]{%
 \ifnum #1\expandafter \@firstoftwo
 \else \expandafter \@secondoftwo
 \fi
}%
\providecommand \@ifx [1]{%
 \ifx #1\expandafter \@firstoftwo
 \else \expandafter \@secondoftwo
 \fi
}%
\providecommand \natexlab [1]{#1}%
\providecommand \enquote  [1]{``#1''}%
\providecommand \bibnamefont  [1]{#1}%
\providecommand \bibfnamefont [1]{#1}%
\providecommand \citenamefont [1]{#1}%
\providecommand \href@noop [0]{\@secondoftwo}%
\providecommand \href [0]{\begingroup \@sanitize@url \@href}%
\providecommand \@href[1]{\@@startlink{#1}\@@href}%
\providecommand \@@href[1]{\endgroup#1\@@endlink}%
\providecommand \@sanitize@url [0]{\catcode `\\12\catcode `\$12\catcode
  `\&12\catcode `\#12\catcode `\^12\catcode `\_12\catcode `\%12\relax}%
\providecommand \@@startlink[1]{}%
\providecommand \@@endlink[0]{}%
\providecommand \url  [0]{\begingroup\@sanitize@url \@url }%
\providecommand \@url [1]{\endgroup\@href {#1}{\urlprefix }}%
\providecommand \urlprefix  [0]{URL }%
\providecommand \Eprint [0]{\href }%
\providecommand \doibase [0]{https://doi.org/}%
\providecommand \selectlanguage [0]{\@gobble}%
\providecommand \bibinfo  [0]{\@secondoftwo}%
\providecommand \bibfield  [0]{\@secondoftwo}%
\providecommand \translation [1]{[#1]}%
\providecommand \BibitemOpen [0]{}%
\providecommand \bibitemStop [0]{}%
\providecommand \bibitemNoStop [0]{.\EOS\space}%
\providecommand \EOS [0]{\spacefactor3000\relax}%
\providecommand \BibitemShut  [1]{\csname bibitem#1\endcsname}%
\let\auto@bib@innerbib\@empty
%</preamble>
\bibitem [{\citenamefont {Peters}(2017)}]{peters2017Reaction}%
  \BibitemOpen
  \bibfield  {author} {\bibinfo {author} {\bibfnamefont {B.}~\bibnamefont
  {Peters}},\ }\href@noop {} {\emph {\bibinfo {title} {Reaction {{Rate Theory}}
  and {{Rare Events}}}}}\ (\bibinfo  {publisher} {Elsevier},\ \bibinfo
  {address} {{Amsterdam}},\ \bibinfo {year} {2017})\BibitemShut {NoStop}%
\bibitem [{\citenamefont {Krivov}(2013)}]{krivov2013Reaction}%
  \BibitemOpen
  \bibfield  {author} {\bibinfo {author} {\bibfnamefont {S.~V.}\ \bibnamefont
  {Krivov}},\ }\bibfield  {title} {\enquote {\bibinfo {title} {On {{Reaction
  Coordinate Optimality}}},}\ }\href {https://doi.org/10.1021/ct3008292}
  {\bibfield  {journal} {\bibinfo  {journal} {J. Chem. Theory Comput.}\
  }\textbf {\bibinfo {volume} {9}},\ \bibinfo {pages} {135--146} (\bibinfo
  {year} {2013})}\BibitemShut {NoStop}%
\bibitem [{\citenamefont {Li}\ and\ \citenamefont {Ma}(2014)}]{li2014Recent}%
  \BibitemOpen
  \bibfield  {author} {\bibinfo {author} {\bibfnamefont {W.}~\bibnamefont
  {Li}}\ and\ \bibinfo {author} {\bibfnamefont {A.}~\bibnamefont {Ma}},\
  }\bibfield  {title} {\enquote {\bibinfo {title} {Recent {{Developments}} in
  {{Methods}} for {{Identifying Reaction Coordinates}}},}\ }\href
  {https://doi.org/10.1080/08927022.2014.907898} {\bibfield  {journal}
  {\bibinfo  {journal} {Mol. Simul.}\ }\textbf {\bibinfo {volume} {40}},\
  \bibinfo {pages} {784--793} (\bibinfo {year} {2014})}\BibitemShut {NoStop}%
\bibitem [{\citenamefont {Peters}\ \emph {et~al.}(2013)\citenamefont {Peters},
  \citenamefont {Bolhuis}, \citenamefont {Mullen},\ and\ \citenamefont
  {Shea}}]{peters2013Reaction}%
  \BibitemOpen
  \bibfield  {author} {\bibinfo {author} {\bibfnamefont {B.}~\bibnamefont
  {Peters}}, \bibinfo {author} {\bibfnamefont {P.~G.}\ \bibnamefont {Bolhuis}},
  \bibinfo {author} {\bibfnamefont {R.~G.}\ \bibnamefont {Mullen}},\ and\
  \bibinfo {author} {\bibfnamefont {J.-E.}\ \bibnamefont {Shea}},\ }\bibfield
  {title} {\enquote {\bibinfo {title} {Reaction coordinates, one-dimensional
  {{Smoluchowski}} equations, and a test for dynamical self-consistency},}\
  }\href {https://doi.org/10.1063/1.4775807} {\bibfield  {journal} {\bibinfo
  {journal} {J. Chem. Phys.}\ }\textbf {\bibinfo {volume} {138}},\ \bibinfo
  {pages} {054106} (\bibinfo {year} {2013})}\BibitemShut {NoStop}%
\bibitem [{\citenamefont {Wales}(2015)}]{wales2015Perspective}%
  \BibitemOpen
  \bibfield  {author} {\bibinfo {author} {\bibfnamefont {D.~J.}\ \bibnamefont
  {Wales}},\ }\bibfield  {title} {\enquote {\bibinfo {title} {Perspective:
  {{Insight}} into reaction coordinates and dynamics from the potential energy
  landscape},}\ }\href {https://doi.org/10.1063/1.4916307} {\bibfield
  {journal} {\bibinfo  {journal} {J. Chem. Phys.}\ }\textbf {\bibinfo {volume}
  {142}},\ \bibinfo {pages} {130901} (\bibinfo {year} {2015})}\BibitemShut
  {NoStop}%
\bibitem [{\citenamefont {Peters}(2016)}]{peters2016Reaction}%
  \BibitemOpen
  \bibfield  {author} {\bibinfo {author} {\bibfnamefont {B.}~\bibnamefont
  {Peters}},\ }\bibfield  {title} {\enquote {\bibinfo {title} {Reaction
  {{Coordinates}} and {{Mechanistic Hypothesis Tests}}},}\ }\href
  {https://doi.org/10.1146/annurev-physchem-040215-112215} {\bibfield
  {journal} {\bibinfo  {journal} {Annu. Rev. Phys. Chem.}\ }\textbf {\bibinfo
  {volume} {67}},\ \bibinfo {pages} {669--690} (\bibinfo {year}
  {2016})}\BibitemShut {NoStop}%
\bibitem [{\citenamefont {Banushkina}\ and\ \citenamefont
  {Krivov}(2016)}]{banushkina2016Optimal}%
  \BibitemOpen
  \bibfield  {author} {\bibinfo {author} {\bibfnamefont {P.~V.}\ \bibnamefont
  {Banushkina}}\ and\ \bibinfo {author} {\bibfnamefont {S.~V.}\ \bibnamefont
  {Krivov}},\ }\bibfield  {title} {\enquote {\bibinfo {title} {Optimal reaction
  coordinates},}\ }\href {https://doi.org/10.1002/wcms.1276} {\bibfield
  {journal} {\bibinfo  {journal} {WIREs Comput. Mol. Sci.}\ }\textbf {\bibinfo
  {volume} {6}},\ \bibinfo {pages} {748--763} (\bibinfo {year}
  {2016})}\BibitemShut {NoStop}%
\bibitem [{\citenamefont {Pietrucci}(2017)}]{pietrucci2017Strategies}%
  \BibitemOpen
  \bibfield  {author} {\bibinfo {author} {\bibfnamefont {F.}~\bibnamefont
  {Pietrucci}},\ }\bibfield  {title} {\enquote {\bibinfo {title} {Strategies
  for the exploration of free energy landscapes: {{Unity}} in diversity and
  challenges ahead},}\ }\href {https://doi.org/10.1016/j.revip.2017.05.001}
  {\bibfield  {journal} {\bibinfo  {journal} {Rev. Phys.}\ }\textbf {\bibinfo
  {volume} {2}},\ \bibinfo {pages} {32--45} (\bibinfo {year}
  {2017})}\BibitemShut {NoStop}%
\bibitem [{\citenamefont {Bolhuis}\ and\ \citenamefont
  {Swenson}(2021)}]{bolhuis2021Transition}%
  \BibitemOpen
  \bibfield  {author} {\bibinfo {author} {\bibfnamefont {P.~G.}\ \bibnamefont
  {Bolhuis}}\ and\ \bibinfo {author} {\bibfnamefont {D.~W.~H.}\ \bibnamefont
  {Swenson}},\ }\bibfield  {title} {\enquote {\bibinfo {title} {Transition
  {{Path Sampling}} as {{Markov Chain Monte Carlo}} of {{Trajectories}}:
  {{Recent Algorithms}}, {{Software}}, {{Applications}}, and {{Future
  Outlook}}},}\ }\href {https://doi.org/10.1002/adts.202000237} {\bibfield
  {journal} {\bibinfo  {journal} {Adv. Theory Simul.}\ }\textbf {\bibinfo
  {volume} {4}},\ \bibinfo {pages} {2000237} (\bibinfo {year}
  {2021})}\BibitemShut {NoStop}%
\bibitem [{\citenamefont {Bolhuis}\ \emph {et~al.}(2002)\citenamefont
  {Bolhuis}, \citenamefont {Chandler}, \citenamefont {Dellago},\ and\
  \citenamefont {Geissler}}]{bolhuis2002Transition}%
  \BibitemOpen
  \bibfield  {author} {\bibinfo {author} {\bibfnamefont {P.~G.}\ \bibnamefont
  {Bolhuis}}, \bibinfo {author} {\bibfnamefont {D.}~\bibnamefont {Chandler}},
  \bibinfo {author} {\bibfnamefont {C.}~\bibnamefont {Dellago}},\ and\ \bibinfo
  {author} {\bibfnamefont {P.~L.}\ \bibnamefont {Geissler}},\ }\bibfield
  {title} {\enquote {\bibinfo {title} {{{Transition Path Sampling}}: {{Throwing
  Ropes Over Rough Mountain Passes}}, in the {{Dark}}},}\ }\href
  {https://doi.org/10.1146/annurev.physchem.53.082301.113146} {\bibfield
  {journal} {\bibinfo  {journal} {Annu. Rev. Phys. Chem.}\ }\textbf {\bibinfo
  {volume} {53}},\ \bibinfo {pages} {291--318} (\bibinfo {year}
  {2002})}\BibitemShut {NoStop}%
\bibitem [{\citenamefont {Rogal}(2021)}]{rogal2021Reaction}%
  \BibitemOpen
  \bibfield  {author} {\bibinfo {author} {\bibfnamefont {J.}~\bibnamefont
  {Rogal}},\ }\bibfield  {title} {\enquote {\bibinfo {title} {Reaction
  coordinates in complex systems-a perspective},}\ }\href
  {https://doi.org/10.1140/epjb/s10051-021-00233-5} {\bibfield  {journal}
  {\bibinfo  {journal} {Eur. Phys. J. B}\ }\textbf {\bibinfo {volume} {94}},\
  \bibinfo {pages} {223} (\bibinfo {year} {2021})}\BibitemShut {NoStop}%
\bibitem [{\citenamefont {Du}\ \emph {et~al.}(1998)\citenamefont {Du},
  \citenamefont {Pande}, \citenamefont {Grosberg}, \citenamefont {Tanaka},\
  and\ \citenamefont {Shakhnovich}}]{du1998Transition}%
  \BibitemOpen
  \bibfield  {author} {\bibinfo {author} {\bibfnamefont {R.}~\bibnamefont
  {Du}}, \bibinfo {author} {\bibfnamefont {V.~S.}\ \bibnamefont {Pande}},
  \bibinfo {author} {\bibfnamefont {A.~Y.}\ \bibnamefont {Grosberg}}, \bibinfo
  {author} {\bibfnamefont {T.}~\bibnamefont {Tanaka}},\ and\ \bibinfo {author}
  {\bibfnamefont {E.~S.}\ \bibnamefont {Shakhnovich}},\ }\bibfield  {title}
  {\enquote {\bibinfo {title} {On the transition coordinate for protein
  folding},}\ }\href {https://doi.org/10.1063/1.475393} {\bibfield  {journal}
  {\bibinfo  {journal} {J. Chem. Phys.}\ }\textbf {\bibinfo {volume} {108}},\
  \bibinfo {pages} {334--350} (\bibinfo {year} {1998})}\BibitemShut {NoStop}%
\bibitem [{\citenamefont {Geissler}, \citenamefont {Dellago},\ and\
  \citenamefont {Chandler}(1999)}]{geissler1999Kinetic}%
  \BibitemOpen
  \bibfield  {author} {\bibinfo {author} {\bibfnamefont {P.~L.}\ \bibnamefont
  {Geissler}}, \bibinfo {author} {\bibfnamefont {C.}~\bibnamefont {Dellago}},\
  and\ \bibinfo {author} {\bibfnamefont {D.}~\bibnamefont {Chandler}},\
  }\bibfield  {title} {\enquote {\bibinfo {title} {Kinetic {{Pathways}} of
  {{Ion Pair Dissociation}} in {{Water}}},}\ }\href
  {https://doi.org/10.1021/jp984837g} {\bibfield  {journal} {\bibinfo
  {journal} {J. Phys. Chem. B}\ }\textbf {\bibinfo {volume} {103}},\ \bibinfo
  {pages} {3706--3710} (\bibinfo {year} {1999})}\BibitemShut {NoStop}%
\bibitem [{\citenamefont {Bolhuis}, \citenamefont {Dellago},\ and\
  \citenamefont {Chandler}(2000)}]{bolhuis2000Reaction}%
  \BibitemOpen
  \bibfield  {author} {\bibinfo {author} {\bibfnamefont {P.~G.}\ \bibnamefont
  {Bolhuis}}, \bibinfo {author} {\bibfnamefont {C.}~\bibnamefont {Dellago}},\
  and\ \bibinfo {author} {\bibfnamefont {D.}~\bibnamefont {Chandler}},\
  }\bibfield  {title} {\enquote {\bibinfo {title} {Reaction coordinates of
  biomolecular isomerization},}\ }\href
  {https://doi.org/10.1073/pnas.100127697} {\bibfield  {journal} {\bibinfo
  {journal} {Proc. Natl. Acad. Sci. U.S.A.}\ }\textbf {\bibinfo {volume}
  {97}},\ \bibinfo {pages} {5877--5882} (\bibinfo {year} {2000})}\BibitemShut
  {NoStop}%
\bibitem [{\citenamefont {Hagan}\ \emph {et~al.}(2003)\citenamefont {Hagan},
  \citenamefont {Dinner}, \citenamefont {Chandler},\ and\ \citenamefont
  {Chakraborty}}]{hagan2003Atomistic}%
  \BibitemOpen
  \bibfield  {author} {\bibinfo {author} {\bibfnamefont {M.~F.}\ \bibnamefont
  {Hagan}}, \bibinfo {author} {\bibfnamefont {A.~R.}\ \bibnamefont {Dinner}},
  \bibinfo {author} {\bibfnamefont {D.}~\bibnamefont {Chandler}},\ and\
  \bibinfo {author} {\bibfnamefont {A.~K.}\ \bibnamefont {Chakraborty}},\
  }\bibfield  {title} {\enquote {\bibinfo {title} {Atomistic understanding of
  kinetic pathways for single base-pair binding and unbinding in {{DNA}}},}\
  }\href {https://doi.org/10.1073/pnas.2036378100} {\bibfield  {journal}
  {\bibinfo  {journal} {Proc. Natl. Acad. Sci. U.S.A.}\ }\textbf {\bibinfo
  {volume} {100}},\ \bibinfo {pages} {13922--13927} (\bibinfo {year}
  {2003})}\BibitemShut {NoStop}%
\bibitem [{\citenamefont {Hummer}(2004)}]{hummer2004Transition}%
  \BibitemOpen
  \bibfield  {author} {\bibinfo {author} {\bibfnamefont {G.}~\bibnamefont
  {Hummer}},\ }\bibfield  {title} {\enquote {\bibinfo {title} {From transition
  paths to transition states and rate coefficients},}\ }\href
  {https://doi.org/10.1063/1.1630572} {\bibfield  {journal} {\bibinfo
  {journal} {J. Chem. Phys.}\ }\textbf {\bibinfo {volume} {120}},\ \bibinfo
  {pages} {516--523} (\bibinfo {year} {2004})}\BibitemShut {NoStop}%
\bibitem [{\citenamefont {Pan}\ and\ \citenamefont
  {Chandler}(2004)}]{pan2004Dynamics}%
  \BibitemOpen
  \bibfield  {author} {\bibinfo {author} {\bibfnamefont {A.~C.}\ \bibnamefont
  {Pan}}\ and\ \bibinfo {author} {\bibfnamefont {D.}~\bibnamefont {Chandler}},\
  }\bibfield  {title} {\enquote {\bibinfo {title} {Dynamics of {{Nucleation}}
  in the {{Ising Model}}},}\ }\href {https://doi.org/10.1021/jp0471249}
  {\bibfield  {journal} {\bibinfo  {journal} {J. Phys. Chem. B}\ }\textbf
  {\bibinfo {volume} {108}},\ \bibinfo {pages} {19681--19686} (\bibinfo {year}
  {2004})}\BibitemShut {NoStop}%
\bibitem [{\citenamefont {Rhee}\ and\ \citenamefont
  {Pande}(2005)}]{rhee2005OneDimensional}%
  \BibitemOpen
  \bibfield  {author} {\bibinfo {author} {\bibfnamefont {Y.~M.}\ \bibnamefont
  {Rhee}}\ and\ \bibinfo {author} {\bibfnamefont {V.~S.}\ \bibnamefont
  {Pande}},\ }\bibfield  {title} {\enquote {\bibinfo {title} {One-{{Dimensional
  Reaction Coordinate}} and the {{Corresponding Potential}} of {{Mean Force}}
  from {{Commitment Probability Distribution}}},}\ }\href
  {https://doi.org/10.1021/jp045544s} {\bibfield  {journal} {\bibinfo
  {journal} {J. Phys. Chem. B}\ }\textbf {\bibinfo {volume} {109}},\ \bibinfo
  {pages} {6780--6786} (\bibinfo {year} {2005})}\BibitemShut {NoStop}%
\bibitem [{\citenamefont {Berezhkovskii}\ and\ \citenamefont
  {Szabo}(2005)}]{berezhkovskii2005Onedimensional}%
  \BibitemOpen
  \bibfield  {author} {\bibinfo {author} {\bibfnamefont {A.}~\bibnamefont
  {Berezhkovskii}}\ and\ \bibinfo {author} {\bibfnamefont {A.}~\bibnamefont
  {Szabo}},\ }\bibfield  {title} {\enquote {\bibinfo {title} {One-dimensional
  reaction coordinates for diffusive activated rate processes in many
  dimensions},}\ }\href {https://doi.org/10.1063/1.1818091} {\bibfield
  {journal} {\bibinfo  {journal} {J. Chem. Phys.}\ }\textbf {\bibinfo {volume}
  {122}},\ \bibinfo {pages} {014503} (\bibinfo {year} {2005})}\BibitemShut
  {NoStop}%
\bibitem [{\citenamefont {Best}\ and\ \citenamefont
  {Hummer}(2005)}]{best2005Reaction}%
  \BibitemOpen
  \bibfield  {author} {\bibinfo {author} {\bibfnamefont {R.~B.}\ \bibnamefont
  {Best}}\ and\ \bibinfo {author} {\bibfnamefont {G.}~\bibnamefont {Hummer}},\
  }\bibfield  {title} {\enquote {\bibinfo {title} {Reaction coordinates and
  rates from transition paths},}\ }\href
  {https://doi.org/10.1073/pnas.0408098102} {\bibfield  {journal} {\bibinfo
  {journal} {Proc. Natl. Acad. Sci. U.S.A.}\ }\textbf {\bibinfo {volume}
  {102}},\ \bibinfo {pages} {6732--6737} (\bibinfo {year} {2005})}\BibitemShut
  {NoStop}%
\bibitem [{\citenamefont {Moroni}, \citenamefont {{ten Wolde}},\ and\
  \citenamefont {Bolhuis}(2005)}]{moroni2005Interplay}%
  \BibitemOpen
  \bibfield  {author} {\bibinfo {author} {\bibfnamefont {D.}~\bibnamefont
  {Moroni}}, \bibinfo {author} {\bibfnamefont {P.~R.}\ \bibnamefont {{ten
  Wolde}}},\ and\ \bibinfo {author} {\bibfnamefont {P.~G.}\ \bibnamefont
  {Bolhuis}},\ }\bibfield  {title} {\enquote {\bibinfo {title} {Interplay
  between {{Structure}} and {{Size}} in a {{Critical Crystal Nucleus}}},}\
  }\href {https://doi.org/10.1103/PhysRevLett.94.235703} {\bibfield  {journal}
  {\bibinfo  {journal} {Phys. Rev. Lett.}\ }\textbf {\bibinfo {volume} {94}},\
  \bibinfo {pages} {235703} (\bibinfo {year} {2005})}\BibitemShut {NoStop}%
\bibitem [{\citenamefont {Peters}(2006)}]{peters2006Using}%
  \BibitemOpen
  \bibfield  {author} {\bibinfo {author} {\bibfnamefont {B.}~\bibnamefont
  {Peters}},\ }\bibfield  {title} {\enquote {\bibinfo {title} {Using the
  histogram test to quantify reaction coordinate error},}\ }\href
  {https://doi.org/10.1063/1.2409924} {\bibfield  {journal} {\bibinfo
  {journal} {J. Chem. Phys.}\ }\textbf {\bibinfo {volume} {125}},\ \bibinfo
  {pages} {241101} (\bibinfo {year} {2006})}\BibitemShut {NoStop}%
\bibitem [{\citenamefont {Branduardi}, \citenamefont {Gervasio},\ and\
  \citenamefont {Parrinello}(2007)}]{branduardi2007Free}%
  \BibitemOpen
  \bibfield  {author} {\bibinfo {author} {\bibfnamefont {D.}~\bibnamefont
  {Branduardi}}, \bibinfo {author} {\bibfnamefont {F.~L.}\ \bibnamefont
  {Gervasio}},\ and\ \bibinfo {author} {\bibfnamefont {M.}~\bibnamefont
  {Parrinello}},\ }\bibfield  {title} {\enquote {\bibinfo {title} {From {{$A$}}
  to {{$B$}} in free energy space},}\ }\href
  {https://doi.org/10.1063/1.2432340} {\bibfield  {journal} {\bibinfo
  {journal} {J. Chem. Phys.}\ }\textbf {\bibinfo {volume} {126}},\ \bibinfo
  {pages} {054103} (\bibinfo {year} {2007})}\BibitemShut {NoStop}%
\bibitem [{\citenamefont {Quaytman}\ and\ \citenamefont
  {Schwartz}(2007)}]{quaytman2007Reaction}%
  \BibitemOpen
  \bibfield  {author} {\bibinfo {author} {\bibfnamefont {S.~L.}\ \bibnamefont
  {Quaytman}}\ and\ \bibinfo {author} {\bibfnamefont {S.~D.}\ \bibnamefont
  {Schwartz}},\ }\bibfield  {title} {\enquote {\bibinfo {title} {Reaction
  coordinate of an enzymatic reaction revealed by transition path sampling},}\
  }\href {https://doi.org/10.1073/pnas.0704304104} {\bibfield  {journal}
  {\bibinfo  {journal} {Proc. Natl. Acad. Sci. U.S.A.}\ }\textbf {\bibinfo
  {volume} {104}},\ \bibinfo {pages} {12253--12258} (\bibinfo {year}
  {2007})}\BibitemShut {NoStop}%
\bibitem [{\citenamefont {Antoniou}\ and\ \citenamefont
  {Schwartz}(2009)}]{antoniou2009Stochastic}%
  \BibitemOpen
  \bibfield  {author} {\bibinfo {author} {\bibfnamefont {D.}~\bibnamefont
  {Antoniou}}\ and\ \bibinfo {author} {\bibfnamefont {S.~D.}\ \bibnamefont
  {Schwartz}},\ }\bibfield  {title} {\enquote {\bibinfo {title} {The stochastic
  separatrix and the reaction coordinate for complex systems},}\ }\href
  {https://doi.org/10.1063/1.3123162} {\bibfield  {journal} {\bibinfo
  {journal} {J. Chem. Phys.}\ }\textbf {\bibinfo {volume} {130}},\ \bibinfo
  {pages} {151103} (\bibinfo {year} {2009})}\BibitemShut {NoStop}%
\bibitem [{\citenamefont {Peters}(2010{\natexlab{a}})}]{peters2010TP}%
  \BibitemOpen
  \bibfield  {author} {\bibinfo {author} {\bibfnamefont {B.}~\bibnamefont
  {Peters}},\ }\bibfield  {title} {\enquote {\bibinfo {title} {P({{TP}}\textbar
  $q$) peak maximization: {{Necessary}} but not sufficient for reaction
  coordinate accuracy},}\ }\href {https://doi.org/10.1016/j.cplett.2010.05.069}
  {\bibfield  {journal} {\bibinfo  {journal} {Chem. Phys. Lett.}\ }\textbf
  {\bibinfo {volume} {494}},\ \bibinfo {pages} {100--103} (\bibinfo {year}
  {2010}{\natexlab{a}})}\BibitemShut {NoStop}%
\bibitem [{\citenamefont {Ernst}, \citenamefont {Wolf},\ and\ \citenamefont
  {Stock}(2017)}]{ernst2017Identification}%
  \BibitemOpen
  \bibfield  {author} {\bibinfo {author} {\bibfnamefont {M.}~\bibnamefont
  {Ernst}}, \bibinfo {author} {\bibfnamefont {S.}~\bibnamefont {Wolf}},\ and\
  \bibinfo {author} {\bibfnamefont {G.}~\bibnamefont {Stock}},\ }\bibfield
  {title} {\enquote {\bibinfo {title} {Identification and {{Validation}} of
  {{Reaction Coordinates Describing Protein Functional Motion}}: {{Hierarchical
  Dynamics}} of {{T4 Lysozyme}}},}\ }\href
  {https://doi.org/10.1021/acs.jctc.7b00571} {\bibfield  {journal} {\bibinfo
  {journal} {J. Chem. Theory Comput.}\ }\textbf {\bibinfo {volume} {13}},\
  \bibinfo {pages} {5076--5088} (\bibinfo {year} {2017})}\BibitemShut {NoStop}%
\bibitem [{\citenamefont {Peters}\ and\ \citenamefont
  {Trout}(2006)}]{peters2006Obtaining}%
  \BibitemOpen
  \bibfield  {author} {\bibinfo {author} {\bibfnamefont {B.}~\bibnamefont
  {Peters}}\ and\ \bibinfo {author} {\bibfnamefont {B.~L.}\ \bibnamefont
  {Trout}},\ }\bibfield  {title} {\enquote {\bibinfo {title} {Obtaining
  reaction coordinates by likelihood maximization},}\ }\href
  {https://doi.org/10.1063/1.2234477} {\bibfield  {journal} {\bibinfo
  {journal} {J. Chem. Phys.}\ }\textbf {\bibinfo {volume} {125}},\ \bibinfo
  {pages} {054108} (\bibinfo {year} {2006})}\BibitemShut {NoStop}%
\bibitem [{\citenamefont {Peters}, \citenamefont {Beckham},\ and\ \citenamefont
  {Trout}(2007)}]{peters2007Extensions}%
  \BibitemOpen
  \bibfield  {author} {\bibinfo {author} {\bibfnamefont {B.}~\bibnamefont
  {Peters}}, \bibinfo {author} {\bibfnamefont {G.~T.}\ \bibnamefont
  {Beckham}},\ and\ \bibinfo {author} {\bibfnamefont {B.~L.}\ \bibnamefont
  {Trout}},\ }\bibfield  {title} {\enquote {\bibinfo {title} {Extensions to the
  likelihood maximization approach for finding reaction coordinates},}\ }\href
  {https://doi.org/10.1063/1.2748396} {\bibfield  {journal} {\bibinfo
  {journal} {J. Chem. Phys.}\ }\textbf {\bibinfo {volume} {127}},\ \bibinfo
  {pages} {034109} (\bibinfo {year} {2007})}\BibitemShut {NoStop}%
\bibitem [{\citenamefont {Peters}(2010{\natexlab{b}})}]{peters2010Recent}%
  \BibitemOpen
  \bibfield  {author} {\bibinfo {author} {\bibfnamefont {B.}~\bibnamefont
  {Peters}},\ }\bibfield  {title} {\enquote {\bibinfo {title} {Recent advances
  in transition path sampling: Accurate reaction coordinates, likelihood
  maximisation and diffusive barrier-crossing dynamics},}\ }\href
  {https://doi.org/10.1080/08927020903536382} {\bibfield  {journal} {\bibinfo
  {journal} {Mol. Simul.}\ }\textbf {\bibinfo {volume} {36}},\ \bibinfo {pages}
  {1265--1281} (\bibinfo {year} {2010}{\natexlab{b}})}\BibitemShut {NoStop}%
\bibitem [{\citenamefont {Beckham}\ \emph {et~al.}(2007)\citenamefont
  {Beckham}, \citenamefont {Peters}, \citenamefont {Starbuck}, \citenamefont
  {Variankaval},\ and\ \citenamefont {Trout}}]{beckham2007SurfaceMediated}%
  \BibitemOpen
  \bibfield  {author} {\bibinfo {author} {\bibfnamefont {G.~T.}\ \bibnamefont
  {Beckham}}, \bibinfo {author} {\bibfnamefont {B.}~\bibnamefont {Peters}},
  \bibinfo {author} {\bibfnamefont {C.}~\bibnamefont {Starbuck}}, \bibinfo
  {author} {\bibfnamefont {N.}~\bibnamefont {Variankaval}},\ and\ \bibinfo
  {author} {\bibfnamefont {B.~L.}\ \bibnamefont {Trout}},\ }\bibfield  {title}
  {\enquote {\bibinfo {title} {Surface-{{Mediated Nucleation}} in the
  {{Solid-State Polymorph Transformation}} of {{Terephthalic Acid}}},}\ }\href
  {https://doi.org/10.1021/ja0687567} {\bibfield  {journal} {\bibinfo
  {journal} {J. Am. Chem. Soc.}\ }\textbf {\bibinfo {volume} {129}},\ \bibinfo
  {pages} {4714--4723} (\bibinfo {year} {2007})}\BibitemShut {NoStop}%
\bibitem [{\citenamefont {Beckham}, \citenamefont {Peters},\ and\ \citenamefont
  {Trout}(2008)}]{beckham2008Evidence}%
  \BibitemOpen
  \bibfield  {author} {\bibinfo {author} {\bibfnamefont {G.~T.}\ \bibnamefont
  {Beckham}}, \bibinfo {author} {\bibfnamefont {B.}~\bibnamefont {Peters}},\
  and\ \bibinfo {author} {\bibfnamefont {B.~L.}\ \bibnamefont {Trout}},\
  }\bibfield  {title} {\enquote {\bibinfo {title} {Evidence for a {{Size
  Dependent Nucleation Mechanism}} in {{Solid State Polymorph
  Transformations}}},}\ }\href {https://doi.org/10.1021/jp710192u} {\bibfield
  {journal} {\bibinfo  {journal} {J. Phys. Chem. B}\ }\textbf {\bibinfo
  {volume} {112}},\ \bibinfo {pages} {7460--7466} (\bibinfo {year}
  {2008})}\BibitemShut {NoStop}%
\bibitem [{\citenamefont
  {Peters}(2010{\natexlab{c}})}]{peters2010TransitionState}%
  \BibitemOpen
  \bibfield  {author} {\bibinfo {author} {\bibfnamefont {B.}~\bibnamefont
  {Peters}},\ }\bibfield  {title} {\enquote {\bibinfo {title}
  {Transition-{{State Theory}}, {{Dynamics}}, and {{Narrow Time Scale
  Separation}} in the {{Rate-Promoting Vibrations Model}} of {{Enzyme
  Catalysis}}},}\ }\href {https://doi.org/10.1021/ct100051a} {\bibfield
  {journal} {\bibinfo  {journal} {J. Chem. Theory Comput.}\ }\textbf {\bibinfo
  {volume} {6}},\ \bibinfo {pages} {1447--1454} (\bibinfo {year}
  {2010}{\natexlab{c}})}\BibitemShut {NoStop}%
\bibitem [{\citenamefont {Vreede}, \citenamefont {Juraszek},\ and\
  \citenamefont {Bolhuis}(2010)}]{vreede2010Predicting}%
  \BibitemOpen
  \bibfield  {author} {\bibinfo {author} {\bibfnamefont {J.}~\bibnamefont
  {Vreede}}, \bibinfo {author} {\bibfnamefont {J.}~\bibnamefont {Juraszek}},\
  and\ \bibinfo {author} {\bibfnamefont {P.~G.}\ \bibnamefont {Bolhuis}},\
  }\bibfield  {title} {\enquote {\bibinfo {title} {Predicting the reaction
  coordinates of millisecond light-induced conformational changes in
  photoactive yellow protein},}\ }\href
  {https://doi.org/10.1073/pnas.0908754107} {\bibfield  {journal} {\bibinfo
  {journal} {Proc. Natl. Acad. Sci. U.S.A.}\ }\textbf {\bibinfo {volume}
  {107}},\ \bibinfo {pages} {2397--2402} (\bibinfo {year} {2010})}\BibitemShut
  {NoStop}%
\bibitem [{\citenamefont {Lechner}\ \emph {et~al.}(2010)\citenamefont
  {Lechner}, \citenamefont {Rogal}, \citenamefont {Juraszek}, \citenamefont
  {Ensing},\ and\ \citenamefont {Bolhuis}}]{lechner2010Nonlinear}%
  \BibitemOpen
  \bibfield  {author} {\bibinfo {author} {\bibfnamefont {W.}~\bibnamefont
  {Lechner}}, \bibinfo {author} {\bibfnamefont {J.}~\bibnamefont {Rogal}},
  \bibinfo {author} {\bibfnamefont {J.}~\bibnamefont {Juraszek}}, \bibinfo
  {author} {\bibfnamefont {B.}~\bibnamefont {Ensing}},\ and\ \bibinfo {author}
  {\bibfnamefont {P.~G.}\ \bibnamefont {Bolhuis}},\ }\bibfield  {title}
  {\enquote {\bibinfo {title} {Nonlinear reaction coordinate analysis in the
  reweighted path ensemble},}\ }\href {https://doi.org/10.1063/1.3491818}
  {\bibfield  {journal} {\bibinfo  {journal} {J. Chem. Phys.}\ }\textbf
  {\bibinfo {volume} {133}},\ \bibinfo {pages} {174110} (\bibinfo {year}
  {2010})}\BibitemShut {NoStop}%
\bibitem [{\citenamefont {Pan}\ and\ \citenamefont
  {Ricci}(2010)}]{pan2010Molecular}%
  \BibitemOpen
  \bibfield  {author} {\bibinfo {author} {\bibfnamefont {B.}~\bibnamefont
  {Pan}}\ and\ \bibinfo {author} {\bibfnamefont {M.~S.}\ \bibnamefont
  {Ricci}},\ }\bibfield  {title} {\enquote {\bibinfo {title} {Molecular
  {{Mechanism}} of {{Acid-Catalyzed Hydrolysis}} of {{Peptide Bonds Using}} a
  {{Model Compound}}},}\ }\href {https://doi.org/10.1021/jp905411n} {\bibfield
  {journal} {\bibinfo  {journal} {J. Phys. Chem. B}\ }\textbf {\bibinfo
  {volume} {114}},\ \bibinfo {pages} {4389--4399} (\bibinfo {year}
  {2010})}\BibitemShut {NoStop}%
\bibitem [{\citenamefont {Beckham}\ and\ \citenamefont
  {Peters}(2011)}]{beckham2011Optimizing}%
  \BibitemOpen
  \bibfield  {author} {\bibinfo {author} {\bibfnamefont {G.~T.}\ \bibnamefont
  {Beckham}}\ and\ \bibinfo {author} {\bibfnamefont {B.}~\bibnamefont
  {Peters}},\ }\bibfield  {title} {\enquote {\bibinfo {title} {Optimizing
  {{Nucleus Size Metrics}} for {{Liquid}}\textendash{{Solid Nucleation}} from
  {{Transition Paths}} of {{Near-Nanosecond Duration}}},}\ }\href
  {https://doi.org/10.1021/jz2002887} {\bibfield  {journal} {\bibinfo
  {journal} {J. Phys. Chem. Lett.}\ }\textbf {\bibinfo {volume} {2}},\ \bibinfo
  {pages} {1133--1138} (\bibinfo {year} {2011})}\BibitemShut {NoStop}%
\bibitem [{\citenamefont {Peters}(2012)}]{peters2012Inertial}%
  \BibitemOpen
  \bibfield  {author} {\bibinfo {author} {\bibfnamefont {B.}~\bibnamefont
  {Peters}},\ }\bibfield  {title} {\enquote {\bibinfo {title} {Inertial
  likelihood maximization for reaction coordinates with high transmission
  coefficients},}\ }\href {https://doi.org/10.1016/j.cplett.2012.10.051}
  {\bibfield  {journal} {\bibinfo  {journal} {Chem. Phys. Lett.}\ }\textbf
  {\bibinfo {volume} {554}},\ \bibinfo {pages} {248--253} (\bibinfo {year}
  {2012})}\BibitemShut {NoStop}%
\bibitem [{\citenamefont {Xi}, \citenamefont {Shah},\ and\ \citenamefont
  {Trout}(2013)}]{xi2013Hopping}%
  \BibitemOpen
  \bibfield  {author} {\bibinfo {author} {\bibfnamefont {L.}~\bibnamefont
  {Xi}}, \bibinfo {author} {\bibfnamefont {M.}~\bibnamefont {Shah}},\ and\
  \bibinfo {author} {\bibfnamefont {B.~L.}\ \bibnamefont {Trout}},\ }\bibfield
  {title} {\enquote {\bibinfo {title} {Hopping of {{Water}} in a {{Glassy
  Polymer Studied}} via {{Transition Path Sampling}} and {{Likelihood
  Maximization}}},}\ }\href {https://doi.org/10.1021/jp3099973} {\bibfield
  {journal} {\bibinfo  {journal} {J. Phys. Chem. B}\ }\textbf {\bibinfo
  {volume} {117}},\ \bibinfo {pages} {3634--3647} (\bibinfo {year}
  {2013})}\BibitemShut {NoStop}%
\bibitem [{\citenamefont {Jungblut}, \citenamefont {Singraber},\ and\
  \citenamefont {Dellago}(2013)}]{jungblut2013Optimising}%
  \BibitemOpen
  \bibfield  {author} {\bibinfo {author} {\bibfnamefont {S.}~\bibnamefont
  {Jungblut}}, \bibinfo {author} {\bibfnamefont {A.}~\bibnamefont
  {Singraber}},\ and\ \bibinfo {author} {\bibfnamefont {C.}~\bibnamefont
  {Dellago}},\ }\bibfield  {title} {\enquote {\bibinfo {title} {Optimising
  reaction coordinates for crystallisation by tuning the crystallinity
  definition},}\ }\href {https://doi.org/10.1080/00268976.2013.832820}
  {\bibfield  {journal} {\bibinfo  {journal} {Mol. Phys.}\ }\textbf {\bibinfo
  {volume} {111}},\ \bibinfo {pages} {3527--3533} (\bibinfo {year}
  {2013})}\BibitemShut {NoStop}%
\bibitem [{\citenamefont {Mullen}, \citenamefont {Shea},\ and\ \citenamefont
  {Peters}(2014)}]{mullen2014Transmission}%
  \BibitemOpen
  \bibfield  {author} {\bibinfo {author} {\bibfnamefont {R.~G.}\ \bibnamefont
  {Mullen}}, \bibinfo {author} {\bibfnamefont {J.-E.}\ \bibnamefont {Shea}},\
  and\ \bibinfo {author} {\bibfnamefont {B.}~\bibnamefont {Peters}},\
  }\bibfield  {title} {\enquote {\bibinfo {title} {Transmission
  {{Coefficients}}, {{Committors}}, and {{Solvent Coordinates}} in {{Ion-Pair
  Dissociation}}},}\ }\href {https://doi.org/10.1021/ct4009798} {\bibfield
  {journal} {\bibinfo  {journal} {J. Chem. Theory Comput.}\ }\textbf {\bibinfo
  {volume} {10}},\ \bibinfo {pages} {659--667} (\bibinfo {year}
  {2014})}\BibitemShut {NoStop}%
\bibitem [{\citenamefont {Mullen}, \citenamefont {Shea},\ and\ \citenamefont
  {Peters}(2015)}]{mullen2015Easy}%
  \BibitemOpen
  \bibfield  {author} {\bibinfo {author} {\bibfnamefont {R.~G.}\ \bibnamefont
  {Mullen}}, \bibinfo {author} {\bibfnamefont {J.-E.}\ \bibnamefont {Shea}},\
  and\ \bibinfo {author} {\bibfnamefont {B.}~\bibnamefont {Peters}},\
  }\bibfield  {title} {\enquote {\bibinfo {title} {Easy {{Transition Path
  Sampling Methods}}: {{Flexible-Length Aimless Shooting}} and {{Permutation
  Shooting}}},}\ }\href {https://doi.org/10.1021/acs.jctc.5b00032} {\bibfield
  {journal} {\bibinfo  {journal} {J. Chem. Theory Comput.}\ }\textbf {\bibinfo
  {volume} {11}},\ \bibinfo {pages} {2421--2428} (\bibinfo {year}
  {2015})}\BibitemShut {NoStop}%
\bibitem [{\citenamefont {Lupi}, \citenamefont {Peters},\ and\ \citenamefont
  {Molinero}(2016)}]{lupi2016Preordering}%
  \BibitemOpen
  \bibfield  {author} {\bibinfo {author} {\bibfnamefont {L.}~\bibnamefont
  {Lupi}}, \bibinfo {author} {\bibfnamefont {B.}~\bibnamefont {Peters}},\ and\
  \bibinfo {author} {\bibfnamefont {V.}~\bibnamefont {Molinero}},\ }\bibfield
  {title} {\enquote {\bibinfo {title} {Pre-ordering of interfacial water in the
  pathway of heterogeneous ice nucleation does not lead to a two-step
  crystallization mechanism},}\ }\href {https://doi.org/10.1063/1.4961652}
  {\bibfield  {journal} {\bibinfo  {journal} {J. Chem. Phys.}\ }\textbf
  {\bibinfo {volume} {145}},\ \bibinfo {pages} {211910} (\bibinfo {year}
  {2016})}\BibitemShut {NoStop}%
\bibitem [{\citenamefont {Jung}, \citenamefont {Okazaki},\ and\ \citenamefont
  {Hummer}(2017)}]{jung2017Transition}%
  \BibitemOpen
  \bibfield  {author} {\bibinfo {author} {\bibfnamefont {H.}~\bibnamefont
  {Jung}}, \bibinfo {author} {\bibfnamefont {K.-i.}\ \bibnamefont {Okazaki}},\
  and\ \bibinfo {author} {\bibfnamefont {G.}~\bibnamefont {Hummer}},\
  }\bibfield  {title} {\enquote {\bibinfo {title} {Transition path sampling of
  rare events by shooting from the top},}\ }\href
  {https://doi.org/10.1063/1.4997378} {\bibfield  {journal} {\bibinfo
  {journal} {J. Chem. Phys.}\ }\textbf {\bibinfo {volume} {147}},\ \bibinfo
  {pages} {152716} (\bibinfo {year} {2017})}\BibitemShut {NoStop}%
\bibitem [{\citenamefont {Joswiak}, \citenamefont {Doherty},\ and\
  \citenamefont {Peters}(2018)}]{joswiak2018Ion}%
  \BibitemOpen
  \bibfield  {author} {\bibinfo {author} {\bibfnamefont {M.~N.}\ \bibnamefont
  {Joswiak}}, \bibinfo {author} {\bibfnamefont {M.~F.}\ \bibnamefont
  {Doherty}},\ and\ \bibinfo {author} {\bibfnamefont {B.}~\bibnamefont
  {Peters}},\ }\bibfield  {title} {\enquote {\bibinfo {title} {Ion dissolution
  mechanism and kinetics at kink sites on {{NaCl}} surfaces},}\ }\href
  {https://doi.org/10.1073/pnas.1713452115} {\bibfield  {journal} {\bibinfo
  {journal} {Proc. Natl. Acad. Sci. U.S.A.}\ }\textbf {\bibinfo {volume}
  {115}},\ \bibinfo {pages} {656--661} (\bibinfo {year} {2018})}\BibitemShut
  {NoStop}%
\bibitem [{\citenamefont {D{\'i}az~Leines}\ and\ \citenamefont
  {Rogal}(2018)}]{diazleines2018Maximum}%
  \BibitemOpen
  \bibfield  {author} {\bibinfo {author} {\bibfnamefont {G.}~\bibnamefont
  {D{\'i}az~Leines}}\ and\ \bibinfo {author} {\bibfnamefont {J.}~\bibnamefont
  {Rogal}},\ }\bibfield  {title} {\enquote {\bibinfo {title} {Maximum
  {{Likelihood Analysis}} of {{Reaction Coordinates}} during {{Solidification}}
  in {{Ni}}},}\ }\href {https://doi.org/10.1021/acs.jpcb.8b08718} {\bibfield
  {journal} {\bibinfo  {journal} {J. Phys. Chem. B}\ }\textbf {\bibinfo
  {volume} {122}},\ \bibinfo {pages} {10934--10942} (\bibinfo {year}
  {2018})}\BibitemShut {NoStop}%
\bibitem [{\citenamefont {Okazaki}\ \emph {et~al.}(2019)\citenamefont
  {Okazaki}, \citenamefont {W{\"o}hlert}, \citenamefont {Warnau}, \citenamefont
  {Jung}, \citenamefont {Yildiz}, \citenamefont {K{\"u}hlbrandt},\ and\
  \citenamefont {Hummer}}]{okazaki2019Mechanism}%
  \BibitemOpen
  \bibfield  {author} {\bibinfo {author} {\bibfnamefont {K.-i.}\ \bibnamefont
  {Okazaki}}, \bibinfo {author} {\bibfnamefont {D.}~\bibnamefont
  {W{\"o}hlert}}, \bibinfo {author} {\bibfnamefont {J.}~\bibnamefont {Warnau}},
  \bibinfo {author} {\bibfnamefont {H.}~\bibnamefont {Jung}}, \bibinfo {author}
  {\bibfnamefont {{\"O}.}~\bibnamefont {Yildiz}}, \bibinfo {author}
  {\bibfnamefont {W.}~\bibnamefont {K{\"u}hlbrandt}},\ and\ \bibinfo {author}
  {\bibfnamefont {G.}~\bibnamefont {Hummer}},\ }\bibfield  {title} {\enquote
  {\bibinfo {title} {Mechanism of the electroneutral sodium/proton antiporter
  {{PaNhaP}} from transition-path shooting},}\ }\href
  {https://doi.org/10.1038/s41467-019-09739-0} {\bibfield  {journal} {\bibinfo
  {journal} {Nat. Commun.}\ }\textbf {\bibinfo {volume} {10}},\ \bibinfo
  {pages} {1742} (\bibinfo {year} {2019})}\BibitemShut {NoStop}%
\bibitem [{\citenamefont {{Arjun}}, \citenamefont {Berendsen},\ and\
  \citenamefont {Bolhuis}(2019)}]{arjun2019Unbiased}%
  \BibitemOpen
  \bibfield  {author} {\bibinfo {author} {\bibnamefont {{Arjun}}}, \bibinfo
  {author} {\bibfnamefont {T.~A.}\ \bibnamefont {Berendsen}},\ and\ \bibinfo
  {author} {\bibfnamefont {P.~G.}\ \bibnamefont {Bolhuis}},\ }\bibfield
  {title} {\enquote {\bibinfo {title} {Unbiased atomistic insight in the
  competing nucleation mechanisms of methane hydrates},}\ }\href
  {https://doi.org/10.1073/pnas.1906502116} {\bibfield  {journal} {\bibinfo
  {journal} {Proc. Natl. Acad. Sci. USA}\ }\textbf {\bibinfo {volume} {116}},\
  \bibinfo {pages} {19305--19310} (\bibinfo {year} {2019})}\BibitemShut
  {NoStop}%
\bibitem [{\citenamefont {Liang}\ \emph {et~al.}(2020)\citenamefont {Liang},
  \citenamefont {D{\'i}az~Leines}, \citenamefont {Drautz},\ and\ \citenamefont
  {Rogal}}]{liang2020Identification}%
  \BibitemOpen
  \bibfield  {author} {\bibinfo {author} {\bibfnamefont {Y.}~\bibnamefont
  {Liang}}, \bibinfo {author} {\bibfnamefont {G.}~\bibnamefont
  {D{\'i}az~Leines}}, \bibinfo {author} {\bibfnamefont {R.}~\bibnamefont
  {Drautz}},\ and\ \bibinfo {author} {\bibfnamefont {J.}~\bibnamefont
  {Rogal}},\ }\bibfield  {title} {\enquote {\bibinfo {title} {Identification of
  a multi-dimensional reaction coordinate for crystal nucleation in
  {{Ni}}{\textsubscript{3}}{{Al}}},}\ }\href
  {https://doi.org/10.1063/5.0010074} {\bibfield  {journal} {\bibinfo
  {journal} {J. Chem. Phys.}\ }\textbf {\bibinfo {volume} {152}},\ \bibinfo
  {pages} {224504} (\bibinfo {year} {2020})}\BibitemShut {NoStop}%
\bibitem [{\citenamefont {Rogers}\ and\ \citenamefont
  {Geissler}(2020)}]{rogers2020Breakage}%
  \BibitemOpen
  \bibfield  {author} {\bibinfo {author} {\bibfnamefont {J.~R.}\ \bibnamefont
  {Rogers}}\ and\ \bibinfo {author} {\bibfnamefont {P.~L.}\ \bibnamefont
  {Geissler}},\ }\bibfield  {title} {\enquote {\bibinfo {title} {Breakage of
  {{Hydrophobic Contacts Limits}} the {{Rate}} of {{Passive Lipid Exchange}}
  between {{Membranes}}},}\ }\href {https://doi.org/10.1021/acs.jpcb.0c04139}
  {\bibfield  {journal} {\bibinfo  {journal} {J. Phys. Chem. B}\ }\textbf
  {\bibinfo {volume} {124}},\ \bibinfo {pages} {5884--5898} (\bibinfo {year}
  {2020})}\BibitemShut {NoStop}%
\bibitem [{\citenamefont {Schwierz}(2020)}]{schwierz2020Kinetic}%
  \BibitemOpen
  \bibfield  {author} {\bibinfo {author} {\bibfnamefont {N.}~\bibnamefont
  {Schwierz}},\ }\bibfield  {title} {\enquote {\bibinfo {title} {Kinetic
  pathways of water exchange in the first hydration shell of magnesium},}\
  }\href {https://doi.org/10.1063/1.5144258} {\bibfield  {journal} {\bibinfo
  {journal} {J. Chem. Phys.}\ }\textbf {\bibinfo {volume} {152}},\ \bibinfo
  {pages} {224106} (\bibinfo {year} {2020})}\BibitemShut {NoStop}%
\bibitem [{\citenamefont {Levintov}, \citenamefont {Paul},\ and\ \citenamefont
  {Vashisth}(2021)}]{levintov2021Reaction}%
  \BibitemOpen
  \bibfield  {author} {\bibinfo {author} {\bibfnamefont {L.}~\bibnamefont
  {Levintov}}, \bibinfo {author} {\bibfnamefont {S.}~\bibnamefont {Paul}},\
  and\ \bibinfo {author} {\bibfnamefont {H.}~\bibnamefont {Vashisth}},\
  }\bibfield  {title} {\enquote {\bibinfo {title} {Reaction {{Coordinate}} and
  {{Thermodynamics}} of {{Base Flipping}} in {{RNA}}},}\ }\href
  {https://doi.org/10.1021/acs.jctc.0c01199} {\bibfield  {journal} {\bibinfo
  {journal} {J. Chem. Theory Comput.}\ }\textbf {\bibinfo {volume} {17}},\
  \bibinfo {pages} {1914--1921} (\bibinfo {year} {2021})}\BibitemShut {NoStop}%
\bibitem [{\citenamefont {Silveira}\ \emph {et~al.}(2021)\citenamefont
  {Silveira}, \citenamefont {Knott}, \citenamefont {Pereira}, \citenamefont
  {Crowley}, \citenamefont {Skaf},\ and\ \citenamefont
  {Beckham}}]{silveira2021Transitiona}%
  \BibitemOpen
  \bibfield  {author} {\bibinfo {author} {\bibfnamefont {R.~L.}\ \bibnamefont
  {Silveira}}, \bibinfo {author} {\bibfnamefont {B.~C.}\ \bibnamefont {Knott}},
  \bibinfo {author} {\bibfnamefont {C.~S.}\ \bibnamefont {Pereira}}, \bibinfo
  {author} {\bibfnamefont {M.~F.}\ \bibnamefont {Crowley}}, \bibinfo {author}
  {\bibfnamefont {M.~S.}\ \bibnamefont {Skaf}},\ and\ \bibinfo {author}
  {\bibfnamefont {G.~T.}\ \bibnamefont {Beckham}},\ }\bibfield  {title}
  {\enquote {\bibinfo {title} {Transition {{Path Sampling Study}} of the
  {{Feruloyl Esterase Mechanism}}},}\ }\href
  {https://doi.org/10.1021/acs.jpcb.0c09725} {\bibfield  {journal} {\bibinfo
  {journal} {J. Phys. Chem. B}\ }\textbf {\bibinfo {volume} {125}},\ \bibinfo
  {pages} {2018--2030} (\bibinfo {year} {2021})}\BibitemShut {NoStop}%
\bibitem [{\citenamefont {Mori}\ and\ \citenamefont
  {Saito}(2020)}]{mori2020Dissecting}%
  \BibitemOpen
  \bibfield  {author} {\bibinfo {author} {\bibfnamefont {T.}~\bibnamefont
  {Mori}}\ and\ \bibinfo {author} {\bibfnamefont {S.}~\bibnamefont {Saito}},\
  }\bibfield  {title} {\enquote {\bibinfo {title} {Dissecting the {{Dynamics}}
  during {{Enzyme Catalysis}}: A {{Case Study}} of {{Pin1 Peptidyl}}-{{Prolyl
  Isomerase}}},}\ }\href {https://doi.org/10.1021/acs.jctc.9b01279} {\bibfield
  {journal} {\bibinfo  {journal} {J. Chem. Theory Comput.}\ }\textbf {\bibinfo
  {volume} {16}},\ \bibinfo {pages} {3396--3407} (\bibinfo {year}
  {2020})}\BibitemShut {NoStop}%
\bibitem [{\citenamefont {Mori}\ \emph {et~al.}(2020)\citenamefont {Mori},
  \citenamefont {Okazaki}, \citenamefont {Mori}, \citenamefont {Kim},\ and\
  \citenamefont {Matubayasi}}]{mori2020Learning}%
  \BibitemOpen
  \bibfield  {author} {\bibinfo {author} {\bibfnamefont {Y.}~\bibnamefont
  {Mori}}, \bibinfo {author} {\bibfnamefont {K.-i.}\ \bibnamefont {Okazaki}},
  \bibinfo {author} {\bibfnamefont {T.}~\bibnamefont {Mori}}, \bibinfo {author}
  {\bibfnamefont {K.}~\bibnamefont {Kim}},\ and\ \bibinfo {author}
  {\bibfnamefont {N.}~\bibnamefont {Matubayasi}},\ }\bibfield  {title}
  {\enquote {\bibinfo {title} {Learning reaction coordinates via cross-entropy
  minimization: Application to alanine dipeptide},}\ }\href
  {https://doi.org/10.1063/5.0009066} {\bibfield  {journal} {\bibinfo
  {journal} {J. Chem. Phys.}\ }\textbf {\bibinfo {volume} {153}},\ \bibinfo
  {pages} {054115} (\bibinfo {year} {2020})}\BibitemShut {NoStop}%
\bibitem [{\citenamefont {Ma}\ and\ \citenamefont
  {Dinner}(2005)}]{ma2005Automatic}%
  \BibitemOpen
  \bibfield  {author} {\bibinfo {author} {\bibfnamefont {A.}~\bibnamefont
  {Ma}}\ and\ \bibinfo {author} {\bibfnamefont {A.~R.}\ \bibnamefont
  {Dinner}},\ }\bibfield  {title} {\enquote {\bibinfo {title} {Automatic
  {{Method}} for {{Identifying Reaction Coordinates}} in {{Complex
  Systems}}},}\ }\href {https://doi.org/10.1021/jp045546c} {\bibfield
  {journal} {\bibinfo  {journal} {J. Phys. Chem. B}\ }\textbf {\bibinfo
  {volume} {109}},\ \bibinfo {pages} {6769--6779} (\bibinfo {year}
  {2005})}\BibitemShut {NoStop}%
\bibitem [{\citenamefont {Sultan}\ and\ \citenamefont
  {Pande}(2018)}]{sultan2018Automated}%
  \BibitemOpen
  \bibfield  {author} {\bibinfo {author} {\bibfnamefont {M.~M.}\ \bibnamefont
  {Sultan}}\ and\ \bibinfo {author} {\bibfnamefont {V.~S.}\ \bibnamefont
  {Pande}},\ }\bibfield  {title} {\enquote {\bibinfo {title} {Automated design
  of collective variables using supervised machine learning},}\ }\href
  {https://doi.org/10.1063/1.5029972} {\bibfield  {journal} {\bibinfo
  {journal} {J. Chem. Phys.}\ }\textbf {\bibinfo {volume} {149}},\ \bibinfo
  {pages} {094106} (\bibinfo {year} {2018})}\BibitemShut {NoStop}%
\bibitem [{\citenamefont {Wehmeyer}\ and\ \citenamefont
  {No{\'e}}(2018)}]{wehmeyer2018Timelagged}%
  \BibitemOpen
  \bibfield  {author} {\bibinfo {author} {\bibfnamefont {C.}~\bibnamefont
  {Wehmeyer}}\ and\ \bibinfo {author} {\bibfnamefont {F.}~\bibnamefont
  {No{\'e}}},\ }\bibfield  {title} {\enquote {\bibinfo {title} {Time-lagged
  autoencoders: {{Deep}} learning of slow collective variables for molecular
  kinetics},}\ }\href {https://doi.org/10.1063/1.5011399} {\bibfield  {journal}
  {\bibinfo  {journal} {J. Chem. Phys.}\ }\textbf {\bibinfo {volume} {148}},\
  \bibinfo {pages} {241703} (\bibinfo {year} {2018})}\BibitemShut {NoStop}%
\bibitem [{\citenamefont {Mardt}\ \emph {et~al.}(2018)\citenamefont {Mardt},
  \citenamefont {Pasquali}, \citenamefont {Wu},\ and\ \citenamefont
  {No{\'e}}}]{mardt2018VAMPnets}%
  \BibitemOpen
  \bibfield  {author} {\bibinfo {author} {\bibfnamefont {A.}~\bibnamefont
  {Mardt}}, \bibinfo {author} {\bibfnamefont {L.}~\bibnamefont {Pasquali}},
  \bibinfo {author} {\bibfnamefont {H.}~\bibnamefont {Wu}},\ and\ \bibinfo
  {author} {\bibfnamefont {F.}~\bibnamefont {No{\'e}}},\ }\bibfield  {title}
  {\enquote {\bibinfo {title} {{{VAMPnets}} for deep learning of molecular
  kinetics},}\ }\href {https://doi.org/10.1038/s41467-017-02388-1} {\bibfield
  {journal} {\bibinfo  {journal} {Nat. Commun.}\ }\textbf {\bibinfo {volume}
  {9}},\ \bibinfo {pages} {5} (\bibinfo {year} {2018})}\BibitemShut {NoStop}%
\bibitem [{\citenamefont {Bittracher}, \citenamefont {Banisch},\ and\
  \citenamefont {Sch{\"u}tte}(2018)}]{bittracher2018Datadriven}%
  \BibitemOpen
  \bibfield  {author} {\bibinfo {author} {\bibfnamefont {A.}~\bibnamefont
  {Bittracher}}, \bibinfo {author} {\bibfnamefont {R.}~\bibnamefont
  {Banisch}},\ and\ \bibinfo {author} {\bibfnamefont {C.}~\bibnamefont
  {Sch{\"u}tte}},\ }\bibfield  {title} {\enquote {\bibinfo {title} {Data-driven
  computation of molecular reaction coordinates},}\ }\href
  {https://doi.org/10.1063/1.5035183} {\bibfield  {journal} {\bibinfo
  {journal} {J. Chem. Phys.}\ }\textbf {\bibinfo {volume} {149}},\ \bibinfo
  {pages} {154103} (\bibinfo {year} {2018})}\BibitemShut {NoStop}%
\bibitem [{\citenamefont {Chen}\ and\ \citenamefont
  {Ferguson}(2018)}]{chen2018Molecular}%
  \BibitemOpen
  \bibfield  {author} {\bibinfo {author} {\bibfnamefont {W.}~\bibnamefont
  {Chen}}\ and\ \bibinfo {author} {\bibfnamefont {A.~L.}\ \bibnamefont
  {Ferguson}},\ }\bibfield  {title} {\enquote {\bibinfo {title} {Molecular
  enhanced sampling with autoencoders: {{On}}-the-fly collective variable
  discovery and accelerated free energy landscape exploration},}\ }\href
  {https://doi.org/10.1002/jcc.25520} {\bibfield  {journal} {\bibinfo
  {journal} {J. Comput. Chem.}\ }\textbf {\bibinfo {volume} {39}},\ \bibinfo
  {pages} {2079--2102} (\bibinfo {year} {2018})}\BibitemShut {NoStop}%
\bibitem [{\citenamefont {Ribeiro}\ \emph {et~al.}(2018)\citenamefont
  {Ribeiro}, \citenamefont {Bravo}, \citenamefont {Wang},\ and\ \citenamefont
  {Tiwary}}]{ribeiro2018Reweighted}%
  \BibitemOpen
  \bibfield  {author} {\bibinfo {author} {\bibfnamefont {J.~M.~L.}\
  \bibnamefont {Ribeiro}}, \bibinfo {author} {\bibfnamefont {P.}~\bibnamefont
  {Bravo}}, \bibinfo {author} {\bibfnamefont {Y.}~\bibnamefont {Wang}},\ and\
  \bibinfo {author} {\bibfnamefont {P.}~\bibnamefont {Tiwary}},\ }\bibfield
  {title} {\enquote {\bibinfo {title} {Reweighted autoencoded variational
  {{Bayes}} for enhanced sampling ({{RAVE}})},}\ }\href
  {https://doi.org/10.1063/1.5025487} {\bibfield  {journal} {\bibinfo
  {journal} {J. Chem. Phys.}\ }\textbf {\bibinfo {volume} {149}},\ \bibinfo
  {pages} {072301} (\bibinfo {year} {2018})}\BibitemShut {NoStop}%
\bibitem [{\citenamefont {Rogal}, \citenamefont {Schneider},\ and\
  \citenamefont {Tuckerman}(2019)}]{rogal2019NeuralNetworkBased}%
  \BibitemOpen
  \bibfield  {author} {\bibinfo {author} {\bibfnamefont {J.}~\bibnamefont
  {Rogal}}, \bibinfo {author} {\bibfnamefont {E.}~\bibnamefont {Schneider}},\
  and\ \bibinfo {author} {\bibfnamefont {M.~E.}\ \bibnamefont {Tuckerman}},\
  }\bibfield  {title} {\enquote {\bibinfo {title} {Neural-{{Network-Based Path
  Collective Variables}} for {{Enhanced Sampling}} of {{Phase
  Transformations}}},}\ }\href {https://doi.org/10.1103/PhysRevLett.123.245701}
  {\bibfield  {journal} {\bibinfo  {journal} {Phys. Rev. Lett.}\ }\textbf
  {\bibinfo {volume} {123}},\ \bibinfo {pages} {245701} (\bibinfo {year}
  {2019})}\BibitemShut {NoStop}%
\bibitem [{\citenamefont {Bonati}, \citenamefont {Rizzi},\ and\ \citenamefont
  {Parrinello}(2020)}]{bonati2020DataDriven}%
  \BibitemOpen
  \bibfield  {author} {\bibinfo {author} {\bibfnamefont {L.}~\bibnamefont
  {Bonati}}, \bibinfo {author} {\bibfnamefont {V.}~\bibnamefont {Rizzi}},\ and\
  \bibinfo {author} {\bibfnamefont {M.}~\bibnamefont {Parrinello}},\ }\bibfield
   {title} {\enquote {\bibinfo {title} {Data-{{Driven Collective Variables}}
  for {{Enhanced Sampling}}},}\ }\href
  {https://doi.org/10.1021/acs.jpclett.0c00535} {\bibfield  {journal} {\bibinfo
   {journal} {J. Phys. Chem. Lett.}\ }\textbf {\bibinfo {volume} {11}},\
  \bibinfo {pages} {2998--3004} (\bibinfo {year} {2020})}\BibitemShut {NoStop}%
\bibitem [{\citenamefont {Wang}, \citenamefont {Lamim~Ribeiro},\ and\
  \citenamefont {Tiwary}(2020)}]{wang2020Machine}%
  \BibitemOpen
  \bibfield  {author} {\bibinfo {author} {\bibfnamefont {Y.}~\bibnamefont
  {Wang}}, \bibinfo {author} {\bibfnamefont {J.~M.}\ \bibnamefont
  {Lamim~Ribeiro}},\ and\ \bibinfo {author} {\bibfnamefont {P.}~\bibnamefont
  {Tiwary}},\ }\bibfield  {title} {\enquote {\bibinfo {title} {Machine learning
  approaches for analyzing and enhancing molecular dynamics simulations},}\
  }\href {https://doi.org/10.1016/j.sbi.2019.12.016} {\bibfield  {journal}
  {\bibinfo  {journal} {Curr. Opin. Struct. Biol.}\ }\textbf {\bibinfo {volume}
  {61}},\ \bibinfo {pages} {139--145} (\bibinfo {year} {2020})}\BibitemShut
  {NoStop}%
\bibitem [{\citenamefont {Wang}\ and\ \citenamefont
  {Tiwary}(2021)}]{wang2021State}%
  \BibitemOpen
  \bibfield  {author} {\bibinfo {author} {\bibfnamefont {D.}~\bibnamefont
  {Wang}}\ and\ \bibinfo {author} {\bibfnamefont {P.}~\bibnamefont {Tiwary}},\
  }\bibfield  {title} {\enquote {\bibinfo {title} {State predictive information
  bottleneck},}\ }\href {https://doi.org/10.1063/5.0038198} {\bibfield
  {journal} {\bibinfo  {journal} {J. Chem. Phys.}\ }\textbf {\bibinfo {volume}
  {154}},\ \bibinfo {pages} {134111} (\bibinfo {year} {2021})}\BibitemShut
  {NoStop}%
\bibitem [{\citenamefont {Sidky}, \citenamefont {Chen},\ and\ \citenamefont
  {Ferguson}(2020)}]{sidky2020Machine}%
  \BibitemOpen
  \bibfield  {author} {\bibinfo {author} {\bibfnamefont {H.}~\bibnamefont
  {Sidky}}, \bibinfo {author} {\bibfnamefont {W.}~\bibnamefont {Chen}},\ and\
  \bibinfo {author} {\bibfnamefont {A.~L.}\ \bibnamefont {Ferguson}},\
  }\bibfield  {title} {\enquote {\bibinfo {title} {Machine learning for
  collective variable discovery and enhanced sampling in biomolecular
  simulation},}\ }\href {https://doi.org/10.1080/00268976.2020.1737742}
  {\bibfield  {journal} {\bibinfo  {journal} {Mol. Phys.}\ }\textbf {\bibinfo
  {volume} {118}},\ \bibinfo {pages} {e1737742} (\bibinfo {year}
  {2020})}\BibitemShut {NoStop}%
\bibitem [{\citenamefont {Zhang}\ \emph {et~al.}(2021)\citenamefont {Zhang},
  \citenamefont {Lei}, \citenamefont {Zhang}, \citenamefont {Han},
  \citenamefont {Li}, \citenamefont {Yang}, \citenamefont {Yang},\ and\
  \citenamefont {Gao}}]{zhang2021Deep}%
  \BibitemOpen
  \bibfield  {author} {\bibinfo {author} {\bibfnamefont {J.}~\bibnamefont
  {Zhang}}, \bibinfo {author} {\bibfnamefont {Y.-K.}\ \bibnamefont {Lei}},
  \bibinfo {author} {\bibfnamefont {Z.}~\bibnamefont {Zhang}}, \bibinfo
  {author} {\bibfnamefont {X.}~\bibnamefont {Han}}, \bibinfo {author}
  {\bibfnamefont {M.}~\bibnamefont {Li}}, \bibinfo {author} {\bibfnamefont
  {L.}~\bibnamefont {Yang}}, \bibinfo {author} {\bibfnamefont {Y.~I.}\
  \bibnamefont {Yang}},\ and\ \bibinfo {author} {\bibfnamefont {Y.~Q.}\
  \bibnamefont {Gao}},\ }\bibfield  {title} {\enquote {\bibinfo {title} {Deep
  reinforcement learning of transition states},}\ }\href
  {https://doi.org/10.1039/D0CP06184K} {\bibfield  {journal} {\bibinfo
  {journal} {Phys. Chem. Chem. Phys.}\ }\textbf {\bibinfo {volume} {23}},\
  \bibinfo {pages} {6888--6895} (\bibinfo {year} {2021})}\BibitemShut {NoStop}%
\bibitem [{\citenamefont {Frassek}, \citenamefont {Arjun},\ and\ \citenamefont
  {Bolhuis}(2021)}]{frassek2021Extended}%
  \BibitemOpen
  \bibfield  {author} {\bibinfo {author} {\bibfnamefont {M.}~\bibnamefont
  {Frassek}}, \bibinfo {author} {\bibfnamefont {A.}~\bibnamefont {Arjun}},\
  and\ \bibinfo {author} {\bibfnamefont {P.~G.}\ \bibnamefont {Bolhuis}},\
  }\bibfield  {title} {\enquote {\bibinfo {title} {An extended autoencoder
  model for reaction coordinate discovery in rare event molecular dynamics
  datasets},}\ }\href {https://doi.org/10.1063/5.0058639} {\bibfield  {journal}
  {\bibinfo  {journal} {J. Chem. Phys.}\ }\textbf {\bibinfo {volume} {155}},\
  \bibinfo {pages} {064103} (\bibinfo {year} {2021})}\BibitemShut {NoStop}%
\bibitem [{\citenamefont {Hooft}, \citenamefont {{P{\'e}rez de Alba
  Ort{\'i}z}},\ and\ \citenamefont {Ensing}(2021)}]{hooft2021Discovering}%
  \BibitemOpen
  \bibfield  {author} {\bibinfo {author} {\bibfnamefont {F.}~\bibnamefont
  {Hooft}}, \bibinfo {author} {\bibfnamefont {A.}~\bibnamefont {{P{\'e}rez de
  Alba Ort{\'i}z}}},\ and\ \bibinfo {author} {\bibfnamefont {B.}~\bibnamefont
  {Ensing}},\ }\bibfield  {title} {\enquote {\bibinfo {title} {Discovering
  {{Collective Variables}} of {{Molecular Transitions}} via {{Genetic
  Algorithms}} and {{Neural Networks}}},}\ }\href
  {https://doi.org/10.1021/acs.jctc.0c00981} {\bibfield  {journal} {\bibinfo
  {journal} {J. Chem. Theory Comput.}\ }\textbf {\bibinfo {volume} {17}},\
  \bibinfo {pages} {2294--2306} (\bibinfo {year} {2021})}\BibitemShut {NoStop}%
\bibitem [{\citenamefont {Bonati}, \citenamefont {Piccini},\ and\ \citenamefont
  {Parrinello}(2021)}]{bonati2021Deepa}%
  \BibitemOpen
  \bibfield  {author} {\bibinfo {author} {\bibfnamefont {L.}~\bibnamefont
  {Bonati}}, \bibinfo {author} {\bibfnamefont {G.}~\bibnamefont {Piccini}},\
  and\ \bibinfo {author} {\bibfnamefont {M.}~\bibnamefont {Parrinello}},\
  }\bibfield  {title} {\enquote {\bibinfo {title} {Deep learning the slow modes
  for rare events sampling},}\ }\href {https://doi.org/10.1073/pnas.2113533118}
  {\bibfield  {journal} {\bibinfo  {journal} {Proc. Natl. Acad. Sci. U.S.A.}\
  }\textbf {\bibinfo {volume} {118}},\ \bibinfo {pages} {e2113533118} (\bibinfo
  {year} {2021})}\BibitemShut {NoStop}%
\bibitem [{\citenamefont {Chen}(2021)}]{chen2021Collective}%
  \BibitemOpen
  \bibfield  {author} {\bibinfo {author} {\bibfnamefont {M.}~\bibnamefont
  {Chen}},\ }\bibfield  {title} {\enquote {\bibinfo {title} {Collective
  variable-based enhanced sampling and machine learning},}\ }\href
  {https://doi.org/10.1140/epjb/s10051-021-00220-w} {\bibfield  {journal}
  {\bibinfo  {journal} {Eur. Phys. J. B}\ }\textbf {\bibinfo {volume} {94}},\
  \bibinfo {pages} {211} (\bibinfo {year} {2021})}\BibitemShut {NoStop}%
\bibitem [{\citenamefont {Belkacemi}\ \emph {et~al.}(2022)\citenamefont
  {Belkacemi}, \citenamefont {Gkeka}, \citenamefont {Leli{\`e}vre},\ and\
  \citenamefont {Stoltz}}]{belkacemi2022Chasing}%
  \BibitemOpen
  \bibfield  {author} {\bibinfo {author} {\bibfnamefont {Z.}~\bibnamefont
  {Belkacemi}}, \bibinfo {author} {\bibfnamefont {P.}~\bibnamefont {Gkeka}},
  \bibinfo {author} {\bibfnamefont {T.}~\bibnamefont {Leli{\`e}vre}},\ and\
  \bibinfo {author} {\bibfnamefont {G.}~\bibnamefont {Stoltz}},\ }\bibfield
  {title} {\enquote {\bibinfo {title} {Chasing {{Collective Variables Using
  Autoencoders}} and {{Biased Trajectories}}},}\ }\href
  {https://doi.org/10.1021/acs.jctc.1c00415} {\bibfield  {journal} {\bibinfo
  {journal} {J. Chem. Theory Comput.}\ }\textbf {\bibinfo {volume} {18}},\
  \bibinfo {pages} {59--78} (\bibinfo {year} {2022})}\BibitemShut {NoStop}%
\bibitem [{\citenamefont {Jung}, \citenamefont {Covino},\ and\ \citenamefont
  {Hummer}(2019)}]{jung2019Artificial}%
  \BibitemOpen
  \bibfield  {author} {\bibinfo {author} {\bibfnamefont {H.}~\bibnamefont
  {Jung}}, \bibinfo {author} {\bibfnamefont {R.}~\bibnamefont {Covino}},\ and\
  \bibinfo {author} {\bibfnamefont {G.}~\bibnamefont {Hummer}},\ }\bibfield
  {title} {\enquote {\bibinfo {title} {Artificial {{Intelligence Assists
  Discovery}} of {{Reaction Coordinates}} and {{Mechanisms}} from {{Molecular
  Dynamics Simulations}}},}\ }\href@noop {} {\  (\bibinfo {year} {2019})},\
  \Eprint {https://arxiv.org/abs/1901.04595} {arXiv:1901.04595} \BibitemShut
  {NoStop}%
\bibitem [{\citenamefont {Jung}\ \emph {et~al.}(2021)\citenamefont {Jung},
  \citenamefont {Covino}, \citenamefont {Arjun}, \citenamefont {Bolhuis},\ and\
  \citenamefont {Hummer}}]{jung2021Autonomous}%
  \BibitemOpen
  \bibfield  {author} {\bibinfo {author} {\bibfnamefont {H.}~\bibnamefont
  {Jung}}, \bibinfo {author} {\bibfnamefont {R.}~\bibnamefont {Covino}},
  \bibinfo {author} {\bibfnamefont {A.}~\bibnamefont {Arjun}}, \bibinfo
  {author} {\bibfnamefont {P.~G.}\ \bibnamefont {Bolhuis}},\ and\ \bibinfo
  {author} {\bibfnamefont {G.}~\bibnamefont {Hummer}},\ }\bibfield  {title}
  {\enquote {\bibinfo {title} {Autonomous artificial intelligence discovers
  mechanisms of molecular self-organization in virtual experiments},}\
  }\href@noop {} {\  (\bibinfo {year} {2021})},\ \Eprint
  {https://arxiv.org/abs/2105.06673} {arXiv:2105.06673} \BibitemShut {NoStop}%
\bibitem [{\citenamefont {Neumann}\ and\ \citenamefont
  {Schwierz}(2022)}]{neumann2022Artificial}%
  \BibitemOpen
  \bibfield  {author} {\bibinfo {author} {\bibfnamefont {J.}~\bibnamefont
  {Neumann}}\ and\ \bibinfo {author} {\bibfnamefont {N.}~\bibnamefont
  {Schwierz}},\ }\bibfield  {title} {\enquote {\bibinfo {title} {Artificial
  {{Intelligence Resolves Kinetic Pathways}} of {{Magnesium Binding}} to
  {{RNA}}},}\ }\href {https://doi.org/10.1021/acs.jctc.1c00752} {\bibfield
  {journal} {\bibinfo  {journal} {J. Chem. Theory Comput.}\ }\textbf {\bibinfo
  {volume} {18}},\ \bibinfo {pages} {1202--1212} (\bibinfo {year}
  {2022})}\BibitemShut {NoStop}%
\bibitem [{\citenamefont {Ren}\ \emph {et~al.}(2005)\citenamefont {Ren},
  \citenamefont {{Vanden-Eijnden}}, \citenamefont {Maragakis},\ and\
  \citenamefont {E}}]{ren2005Transition}%
  \BibitemOpen
  \bibfield  {author} {\bibinfo {author} {\bibfnamefont {W.}~\bibnamefont
  {Ren}}, \bibinfo {author} {\bibfnamefont {E.}~\bibnamefont
  {{Vanden-Eijnden}}}, \bibinfo {author} {\bibfnamefont {P.}~\bibnamefont
  {Maragakis}},\ and\ \bibinfo {author} {\bibfnamefont {W.}~\bibnamefont {E}},\
  }\bibfield  {title} {\enquote {\bibinfo {title} {Transition pathways in
  complex systems: {{Application}} of the finite-temperature string method to
  the alanine dipeptide},}\ }\href {https://doi.org/10.1063/1.2013256}
  {\bibfield  {journal} {\bibinfo  {journal} {J. Chem. Phys.}\ }\textbf
  {\bibinfo {volume} {123}},\ \bibinfo {pages} {134109} (\bibinfo {year}
  {2005})}\BibitemShut {NoStop}%
\bibitem [{\citenamefont {Manuchehrfar}\ \emph {et~al.}(2021)\citenamefont
  {Manuchehrfar}, \citenamefont {Li}, \citenamefont {Tian}, \citenamefont
  {Ma},\ and\ \citenamefont {Liang}}]{manuchehrfar2021Exacta}%
  \BibitemOpen
  \bibfield  {author} {\bibinfo {author} {\bibfnamefont {F.}~\bibnamefont
  {Manuchehrfar}}, \bibinfo {author} {\bibfnamefont {H.}~\bibnamefont {Li}},
  \bibinfo {author} {\bibfnamefont {W.}~\bibnamefont {Tian}}, \bibinfo {author}
  {\bibfnamefont {A.}~\bibnamefont {Ma}},\ and\ \bibinfo {author}
  {\bibfnamefont {J.}~\bibnamefont {Liang}},\ }\bibfield  {title} {\enquote
  {\bibinfo {title} {Exact {{Topology}} of the {{Dynamic Probability Surface}}
  of an {{Activated Process}} by {{Persistent Homology}}},}\ }\href
  {https://doi.org/10.1021/acs.jpcb.1c00904} {\bibfield  {journal} {\bibinfo
  {journal} {J. Phys. Chem. B}\ }\textbf {\bibinfo {volume} {125}},\ \bibinfo
  {pages} {4667--4680} (\bibinfo {year} {2021})}\BibitemShut {NoStop}%
\bibitem [{\citenamefont {Antoniou}, \citenamefont {Abolfath},\ and\
  \citenamefont {Schwartz}(2004)}]{antoniou2004Transition}%
  \BibitemOpen
  \bibfield  {author} {\bibinfo {author} {\bibfnamefont {D.}~\bibnamefont
  {Antoniou}}, \bibinfo {author} {\bibfnamefont {M.~R.}\ \bibnamefont
  {Abolfath}},\ and\ \bibinfo {author} {\bibfnamefont {S.~D.}\ \bibnamefont
  {Schwartz}},\ }\bibfield  {title} {\enquote {\bibinfo {title} {Transition
  path sampling study of classical rate-promoting vibrations},}\ }\href
  {https://doi.org/10.1063/1.1782813} {\bibfield  {journal} {\bibinfo
  {journal} {J. Chem. Phys.}\ }\textbf {\bibinfo {volume} {121}},\ \bibinfo
  {pages} {6442--6447} (\bibinfo {year} {2004})}\BibitemShut {NoStop}%
\bibitem [{\citenamefont {Adadi}\ and\ \citenamefont
  {Berrada}(2018)}]{adadi2018Peeking}%
  \BibitemOpen
  \bibfield  {author} {\bibinfo {author} {\bibfnamefont {A.}~\bibnamefont
  {Adadi}}\ and\ \bibinfo {author} {\bibfnamefont {M.}~\bibnamefont
  {Berrada}},\ }\bibfield  {title} {\enquote {\bibinfo {title} {Peeking
  {{Inside}} the {{Black-Box}}: {{A Survey}} on {{Explainable Artificial
  Intelligence}} ({{XAI}})},}\ }\href
  {https://doi.org/10.1109/ACCESS.2018.2870052} {\bibfield  {journal} {\bibinfo
   {journal} {IEEE Access}\ }\textbf {\bibinfo {volume} {6}},\ \bibinfo {pages}
  {52138--52160} (\bibinfo {year} {2018})}\BibitemShut {NoStop}%
\bibitem [{\citenamefont {Molnar}(2020)}]{molnar2020Interpretable}%
  \BibitemOpen
  \bibfield  {author} {\bibinfo {author} {\bibfnamefont {C.}~\bibnamefont
  {Molnar}},\ }\href@noop {} {\emph {\bibinfo {title} {{Interpretable Machine
  Learning}}}}\ (\bibinfo  {publisher} {Lulu.com},\ \bibinfo {address}
  {{Morisville, North Carolina}},\ \bibinfo {year} {2020})\BibitemShut
  {NoStop}%
\bibitem [{\citenamefont {Ribeiro}, \citenamefont {Singh},\ and\ \citenamefont
  {Guestrin}(2016)}]{ribeiro2016Whya}%
  \BibitemOpen
  \bibfield  {author} {\bibinfo {author} {\bibfnamefont {M.~T.}\ \bibnamefont
  {Ribeiro}}, \bibinfo {author} {\bibfnamefont {S.}~\bibnamefont {Singh}},\
  and\ \bibinfo {author} {\bibfnamefont {C.}~\bibnamefont {Guestrin}},\
  }\bibfield  {title} {\enquote {\bibinfo {title} {"{{Why Should I Trust
  You}}?": Explaining the {{Predictions}} of {{Any Classifier}}},}\ }in\ \href
  {https://doi.org/10.1145/2939672.2939778} {\emph {\bibinfo {booktitle}
  {Proceedings of the 22nd {{ACM SIGKDD International Conference}} on
  {{Knowledge Discovery}} and {{Data Mining}}}}}\ (\bibinfo {address} {{San
  Francisco California, U.S.A.}},\ \bibinfo {year} {2016})\ pp.\ \bibinfo
  {pages} {1135--1144}\BibitemShut {NoStop}%
\bibitem [{\citenamefont {Lundberg}\ and\ \citenamefont
  {Lee}(2017)}]{lundberg2017unified}%
  \BibitemOpen
  \bibfield  {author} {\bibinfo {author} {\bibfnamefont {S.~M.}\ \bibnamefont
  {Lundberg}}\ and\ \bibinfo {author} {\bibfnamefont {S.-I.}\ \bibnamefont
  {Lee}},\ }\bibfield  {title} {\enquote {\bibinfo {title} {A unified approach
  to interpreting model predictions},}\ }in\ \href@noop {} {\emph {\bibinfo
  {booktitle} {Proceedings of the 31st international conference on neural
  information processing systems}}}\ (\bibinfo {year} {2017})\ pp.\ \bibinfo
  {pages} {4768--4777}\BibitemShut {NoStop}%
\bibitem [{\citenamefont {Maas}, \citenamefont {Hannun},\ and\ \citenamefont
  {Ng}(2013)}]{maas2013Rectifier}%
  \BibitemOpen
  \bibfield  {author} {\bibinfo {author} {\bibfnamefont {A.~L.}\ \bibnamefont
  {Maas}}, \bibinfo {author} {\bibfnamefont {A.~Y.}\ \bibnamefont {Hannun}},\
  and\ \bibinfo {author} {\bibfnamefont {A.~Y.}\ \bibnamefont {Ng}},\
  }\bibfield  {title} {\enquote {\bibinfo {title} {Rectifier nonlinearities
  improve neural network acoustic models},}\ }\href@noop {} {\bibfield
  {journal} {\bibinfo  {journal} {Proc. ICML}\ }\textbf {\bibinfo {volume}
  {30}},\ \bibinfo {pages} {3} (\bibinfo {year} {2013})}\BibitemShut {NoStop}%
\bibitem [{\citenamefont {Kingma}\ and\ \citenamefont
  {Ba}(2014)}]{kingma2017Adam}%
  \BibitemOpen
  \bibfield  {author} {\bibinfo {author} {\bibfnamefont {D.~P.}\ \bibnamefont
  {Kingma}}\ and\ \bibinfo {author} {\bibfnamefont {J.}~\bibnamefont {Ba}},\
  }\bibfield  {title} {\enquote {\bibinfo {title} {Adam: A {{Method}} for
  {{Stochastic Optimization}}},}\ }\href@noop {} {\  (\bibinfo {year}
  {2014})},\ \Eprint {https://arxiv.org/abs/1412.6980} {arXiv:1412.6980}
  \BibitemShut {NoStop}%
\bibitem [{\citenamefont {Abadi}\ \emph {et~al.}(2016)\citenamefont {Abadi},
  \citenamefont {Agarwal}, \citenamefont {Barham}, \citenamefont {Brevdo},
  \citenamefont {Chen}, \citenamefont {Citro}, \citenamefont {Corrado},
  \citenamefont {Davis}, \citenamefont {Dean}, \citenamefont {Devin},
  \citenamefont {Ghemawat}, \citenamefont {Goodfellow}, \citenamefont {Harp},
  \citenamefont {Irving}, \citenamefont {Isard}, \citenamefont {Jia},
  \citenamefont {Jozefowicz}, \citenamefont {Kaiser}, \citenamefont {Kudlur},
  \citenamefont {Levenberg}, \citenamefont {Mane}, \citenamefont {Monga},
  \citenamefont {Moore}, \citenamefont {Murray}, \citenamefont {Olah},
  \citenamefont {Schuster}, \citenamefont {Shlens}, \citenamefont {Steiner},
  \citenamefont {Sutskever}, \citenamefont {Talwar}, \citenamefont {Tucker},
  \citenamefont {Vanhoucke}, \citenamefont {Vasudevan}, \citenamefont {Viegas},
  \citenamefont {Vinyals}, \citenamefont {Warden}, \citenamefont {Wattenberg},
  \citenamefont {Wicke}, \citenamefont {Yu},\ and\ \citenamefont
  {Zheng}}]{abadi2016TensorFlow}%
  \BibitemOpen
  \bibfield  {author} {\bibinfo {author} {\bibfnamefont {M.}~\bibnamefont
  {Abadi}}, \bibinfo {author} {\bibfnamefont {A.}~\bibnamefont {Agarwal}},
  \bibinfo {author} {\bibfnamefont {P.}~\bibnamefont {Barham}}, \bibinfo
  {author} {\bibfnamefont {E.}~\bibnamefont {Brevdo}}, \bibinfo {author}
  {\bibfnamefont {Z.}~\bibnamefont {Chen}}, \bibinfo {author} {\bibfnamefont
  {C.}~\bibnamefont {Citro}}, \bibinfo {author} {\bibfnamefont {G.~S.}\
  \bibnamefont {Corrado}}, \bibinfo {author} {\bibfnamefont {A.}~\bibnamefont
  {Davis}}, \bibinfo {author} {\bibfnamefont {J.}~\bibnamefont {Dean}},
  \bibinfo {author} {\bibfnamefont {M.}~\bibnamefont {Devin}}, \bibinfo
  {author} {\bibfnamefont {S.}~\bibnamefont {Ghemawat}}, \bibinfo {author}
  {\bibfnamefont {I.}~\bibnamefont {Goodfellow}}, \bibinfo {author}
  {\bibfnamefont {A.}~\bibnamefont {Harp}}, \bibinfo {author} {\bibfnamefont
  {G.}~\bibnamefont {Irving}}, \bibinfo {author} {\bibfnamefont
  {M.}~\bibnamefont {Isard}}, \bibinfo {author} {\bibfnamefont
  {Y.}~\bibnamefont {Jia}}, \bibinfo {author} {\bibfnamefont {R.}~\bibnamefont
  {Jozefowicz}}, \bibinfo {author} {\bibfnamefont {L.}~\bibnamefont {Kaiser}},
  \bibinfo {author} {\bibfnamefont {M.}~\bibnamefont {Kudlur}}, \bibinfo
  {author} {\bibfnamefont {J.}~\bibnamefont {Levenberg}}, \bibinfo {author}
  {\bibfnamefont {D.}~\bibnamefont {Mane}}, \bibinfo {author} {\bibfnamefont
  {R.}~\bibnamefont {Monga}}, \bibinfo {author} {\bibfnamefont
  {S.}~\bibnamefont {Moore}}, \bibinfo {author} {\bibfnamefont
  {D.}~\bibnamefont {Murray}}, \bibinfo {author} {\bibfnamefont
  {C.}~\bibnamefont {Olah}}, \bibinfo {author} {\bibfnamefont {M.}~\bibnamefont
  {Schuster}}, \bibinfo {author} {\bibfnamefont {J.}~\bibnamefont {Shlens}},
  \bibinfo {author} {\bibfnamefont {B.}~\bibnamefont {Steiner}}, \bibinfo
  {author} {\bibfnamefont {I.}~\bibnamefont {Sutskever}}, \bibinfo {author}
  {\bibfnamefont {K.}~\bibnamefont {Talwar}}, \bibinfo {author} {\bibfnamefont
  {P.}~\bibnamefont {Tucker}}, \bibinfo {author} {\bibfnamefont
  {V.}~\bibnamefont {Vanhoucke}}, \bibinfo {author} {\bibfnamefont
  {V.}~\bibnamefont {Vasudevan}}, \bibinfo {author} {\bibfnamefont
  {F.}~\bibnamefont {Viegas}}, \bibinfo {author} {\bibfnamefont
  {O.}~\bibnamefont {Vinyals}}, \bibinfo {author} {\bibfnamefont
  {P.}~\bibnamefont {Warden}}, \bibinfo {author} {\bibfnamefont
  {M.}~\bibnamefont {Wattenberg}}, \bibinfo {author} {\bibfnamefont
  {M.}~\bibnamefont {Wicke}}, \bibinfo {author} {\bibfnamefont
  {Y.}~\bibnamefont {Yu}},\ and\ \bibinfo {author} {\bibfnamefont
  {X.}~\bibnamefont {Zheng}},\ }\bibfield  {title} {\enquote {\bibinfo {title}
  {{{TensorFlow}}: Large-{{Scale Machine Learning}} on {{Heterogeneous
  Distributed Systems}}},}\ }\href@noop {} {\  (\bibinfo {year} {2016})},\
  \Eprint {https://arxiv.org/abs/1603.04467} {arXiv:1603.04467} \BibitemShut
  {NoStop}%
\bibitem [{\citenamefont {Shapley}(1953)}]{shapley195317}%
  \BibitemOpen
  \bibfield  {author} {\bibinfo {author} {\bibfnamefont {L.~S.}\ \bibnamefont
  {Shapley}},\ }\bibfield  {title} {\enquote {\bibinfo {title} {17. {{A Value}}
  for n-{{Person Games}}},}\ }in\ \href
  {https://doi.org/10.1515/9781400881970-018} {\emph {\bibinfo {booktitle}
  {Contributions to the {{Theory}} of {{Games}} ({{AM-28}}), {{Volume II}}}}},\
  \bibinfo {editor} {edited by\ \bibinfo {editor} {\bibfnamefont {H.~W.}\
  \bibnamefont {Kuhn}}\ and\ \bibinfo {editor} {\bibfnamefont {A.~W.}\
  \bibnamefont {Tucker}}}\ (\bibinfo  {publisher} {Princeton University
  Press},\ \bibinfo {year} {1953})\ pp.\ \bibinfo {pages}
  {307--318}\BibitemShut {NoStop}%
\end{thebibliography}
\end{document}